\documentclass[pdflatex,sn-basic,Numbered]{sn-jnl}

\usepackage{amsmath,amssymb,amsfonts}
\usepackage{graphicx}
\begin{document}

\title{Reliability-Aware Refinement of High-Resolution Human Mobility Origin--Destination Tables under Sparsity and Suppression}

\author[1]{\fnm{Yohei} \sur{Shida}}
\equalcont{These authors contributed equally to this work.}
\author[2]{\fnm{Ryosuke} \sur{Susuta}}
\equalcont{These authors contributed equally to this work.}
\author[2]{\fnm{Takuhiro} \sur{Moriyama}}
\author*[3]{\fnm{Jun'ichi} \sur{Ozaki}}\email{ozaki.jun.fe@yokohama-cu.ac.jp}

\affil[1]{\orgdiv{Institute of Systems and Information Engineering}, \orgname{University of Tsukuba}, \orgaddress{\street{1-1-1 Tennodai}, \city{Tsukuba}, \postcode{305-8573}, \state{Ibaraki}, \country{Japan}}}
\affil[2]{\orgname{GEOTRA Co., Ltd.}, \orgaddress{\street{Otemachi One Tower 6F, 1-2-1 Otemachi}, \city{Chiyoda-ku}, \postcode{100-0004}, \state{Tokyo}, \country{Japan}}}
\affil*[3]{\orgdiv{School of Data Science}, \orgname{Yokohama City University}, \orgaddress{\street{22-2 Seto, Kanazawa-ku}, \city{Yokohama}, \postcode{236-0027}, \state{Kanagawa}, \country{Japan}}}

\abstract{High-resolution origin--destination (OD) tables derived from digital mobility traces can preserve fine-scale collective movement structure, but sampling, population expansion, and privacy-motivated low-count suppression create heterogeneous observational support across OD entries and make apparent zeros difficult to interpret. We develop a reliability-aware refinement framework that combines each released OD entry with a time-resolved, data-informed radiation baseline through a closed-form cell-wise Gamma--Poisson update, without smoothing neighboring OD pairs. Applied to a 125 m hourly Tokyo OD product, the method preserves concentrated directional flows while adding lower-intensity, baseline-supported links. The relative contribution of released and model-supported structure varies substantially across OD pairs and departure hours, including among positive-flow entries: released flows are most strongly retained at 7:00, followed by 17:00, whereas the radiation baseline contributes more at 13:00. The inferred characteristic radius broadens the movement range represented by the baseline and exhibits structured variation around major station areas and rail corridors. Tokyo and Fukuoka also show a common increase in data--baseline separation at long distances. Under controlled Poisson thinning, calibrated refinement yields lower realization-averaged mean absolute error than an independently resampled no-refinement table across the tested conditions. This experiment provides internal calibration under imposed degradation, not external validation against true unsuppressed flows. The framework thus provides a transparent way to preserve fine-scale empirical mobility structure while making the cell-wise contribution of structured model support explicit.}


\maketitle

\section{Introduction}

High-resolution origin--destination (OD) tables derived from digital mobility traces provide a compact representation of collective movement across space and time. They are used to characterize urban activity, quantify mobility flows, and provide inputs to downstream models of accessibility, exposure, emergency response, and epidemic spread~\cite{barbosa2018human,chinazzi2020effect,kraemer2020effect}. Their analytical value is greatest when they are spatially and temporally fine grained, because fine grids can preserve station catchments, corridor effects, localized activity centers, and discontinuities that coarser zoning systems average away~\cite{alessandretti2020scales,zhong2025universal}. At the same time, increasing resolution makes individual OD entries more exposed to sampling, population expansion, and privacy-motivated suppression, complicating the interpretation of the released product.

This observation problem is especially important for mobility aggregates derived from passively collected location traces. Mobile phone, GPS, and cellular signalling data can support large-scale OD construction, but the resulting products depend on platform coverage, sampling processes, population scaling, and provider-side post-processing~\cite{iqbal2014development,toole2015path,fekih2021data,gallotti2024distorted}. Such data can also exhibit systematic bias across users, locations, and spatial scales~\cite{ranjan2012call,li2024understanding}. Fine-scale mobility data are highly privacy-sensitive~\cite{de2013unique}, and release procedures often suppress, generalize, or perturb low-count mobility records to reduce disclosure risk~\cite{matet2023adaptative,haydari2022differentially,kohli2024privacy}. These processes create heterogeneous observational support across OD pairs. A zero entry in a released fine-grid OD table is therefore ambiguous: it may represent no mobility, a low but nonzero flow, or a flow affected by disclosure-control suppression. Because low-count suppression depends on the underlying flow intensity, such entries cannot be treated as ordinary additive noise or random missingness in the usual missing-data sense~\cite{rubin1976inference}.

Existing OD estimation and completion methods provide important foundations, but they generally assume different observation settings. Classical OD estimation and matrix updating methods infer demand from link counts, assignment models, marginal totals, or prior demand matrices~\cite{van1980most,spiess1987maximum,cascetta1988unified}. Entropy, cross-entropy, and iterative proportional fitting approaches update OD tables under margin or flow constraints~\cite{wilson2013entropy,xie2010maximum}. More recent OD completion methods use low-rank, tensor, or graph-regularized assumptions to infer missing entries from partially observed OD matrices or historical OD tensors~\cite{dong2021rail,dong2021riemannian}. Recent physics-informed learning approaches have also sought to reconstruct mobility networks from limited observations and transfer them across cities~\cite{yang2026transferable}. In the downstream release setting considered here, however, the analyst receives a provider-side aggregate after sampling, expansion, and suppression. The problem is therefore not simply to infer an unobserved demand matrix or fill missing cells, but to refine an already released product without erasing empirical mobility structure that remains informative.

This setting changes the inferential target. We ask how released OD information and structured model information can be combined at the individual OD-entry level without assuming that either source should dominate everywhere. Taking the released table at face value fixes apparent zeros at zero, even when they may reflect weak support or suppression. Replacing the table wholesale with a model discards sharp empirical structure where released flows depart from generic spatial-interaction structure. A useful refinement should therefore preserve such empirical departures, avoid smoothing across neighboring OD entries, and allow structured model support to contribute selectively.

We implement this idea with a time-resolved, data-informed radiation baseline and a closed-form cell-wise Gamma--Poisson update. The radiation baseline combines purpose-specific opportunity fields, hour-specific activity mixtures, and an origin- and hour-specific characteristic radius \(r_0(i,t)\)~\cite{simini2012universal}. Because quantities entering the baseline, including origin outflows and \(r_0(i,t)\), are derived in part from the released OD product, the baseline should be understood as a structured spatial-interaction reference rather than a data-independent prior. The update is applied independently to each OD entry, without smoothing neighboring entries in the OD matrix. The resulting estimator explicitly blends the released observation and the radiation baseline, while the data weight records their relative contribution to each refined entry; it is an operational mixing coefficient rather than a calibrated probability of measurement accuracy.

We evaluate the framework on a population-expanded and low-count-masked Tokyo OD product with 125 m spatial resolution and hourly departure intervals. The refined estimates retain the high-valued directional features of the released table while adding lower-intensity, baseline-supported links. Across positive-flow OD entries, the balance between the two inputs varies systematically by departure hour, with the strongest data anchoring at 7:00, followed by 17:00, and greater baseline contribution at 13:00. Distance-profile analyses and the inferred \(r_0(i,t)\) fields provide complementary diagnostics of the spatial scales represented by the radiation baseline, while the supplementary Fukuoka analysis in Additional file 1 examines the same framework in a second metropolitan release context. Controlled Poisson thinning evaluates internal calibration under imposed degradation of the released table rather than recovery of an inaccessible ground-truth OD matrix. Together, these analyses show how a high-resolution mobility product can be refined without wholesale model replacement while retaining explicit information about where and when structured model support contributes more strongly.

\section{Data and release setting}
We estimate a refined OD table by anchoring to the observed aggregated OD table (D1), assigning trip purposes using citywide weights from a Person Trip survey (D2), and using grid-level opportunity (Point of Interest) and population fields (D3 and D4) as inputs to the baseline model. D1 is an aggregated OD dataset constructed by KDDI, a major mobile network operator in Japan, from GPS location traces collected from consenting subscribers for the 23 wards of Tokyo on weekdays in December 2023~\cite{KDDICORP85:online}. The released OD tables are provided in stratified form by demographic attributes (sex and age group: 20s, 30s, 40s, 50s, and 60+), excluding individuals aged 19 years and below. In addition, D1 is released with movement modes classified into walking, car, expressway, railway, and stay, where the mode labels are assigned in advance by an internal KDDI algorithm. OD pairs are included if either the origin or the destination lies within Tokyo's 23 special wards and the other endpoint is within 50 km of the 23-ward boundary. A stay-out is defined as remaining at the same location for at least 15 minutes. OD pairs are extracted in 1 h intervals and are not linked across intervals. The released OD tables are hourly on a 125 m grid; the aggregates are scaled to match population ratios; OD cells with counts below 10 after this population-ratio expansion over the 21-day aggregation period are masked; and the resulting OD tables are normalized by the number of weekdays. The masking rule therefore applies to values below $10/21$ in the daily-average representation. Masked entries are unavailable in the released product and are encoded as zero in the analysis table; no individual identifiers are available. D2 is the 6th Tokyo metropolitan area Person Trip survey conducted in 2018 and is used only to compute citywide departure-hour-by-purpose shares that assign trip purposes, with purposes aggregated into work, home, and stay-out~\cite{PT:online}. D3 is a POI (Point of Interest) dataset from Foursquare aggregated to the 125 m grid to define destination opportunity inputs for work and stay-out using a predefined category mapping~\cite{TheFutur61:online}. D4 is gridded residential population from the Population Census 2020 aligned to the 125 m grid and used as the population field for home locations, providing residential mass that supports both departures from home and returns to home; when population is available at 250 m resolution, it is aligned to 125 m by evenly splitting each 250 m cell into four 125 m subcells~\cite{Populati73:online}.

\section{Methods}

\subsection{Released-product refinement problem}

We refine the released high-resolution hourly OD table by preserving observation-supported OD structure while using the radiation baseline to complement weakly supported or masked entries. The procedure has two stages: an extended radiation model constructs the radiation baseline, and a cell-wise Bayesian update combines it with the data OD.

Let $i$ and $j$ index origin and destination grid cells, $t$ index departure-hour bins, and $c$ index an attribute stratum (e.g., sex $\times$ age group). We denote the released OD data by $T^{\mathrm{data}}_{ij\mid t,c}$ and the latent OD flow by $T_{ij\mid t,c}$. As described in the Data, the daily-average masking threshold is $T_{\mathrm{th}}=10/21$.
We encode masked entries as zeros because their released values are unavailable and are only known to fall below the release threshold. This is an operational encoding rather than an assumption that the latent flow is exactly zero, and it allows the same cell-wise update to be applied without introducing a separate provider-specific missing-data model.

\subsection{Extended radiation baseline}

The first stage constructs the extended radiation baseline $T^{\mathrm{rad}}_{ij\mid t,c}$~\cite{simini2012universal,barbosa2018human,du2025review}. This baseline has two components: a destination-choice share from origin $i$ at departure hour $t$, and an attribute-specific outflow from that origin. We first construct the destination-choice share and then scale it by the origin outflow.

We begin with activity-specific attractiveness fields. Let $a$ index \emph{home}, \emph{work}, and \emph{stay-out}, and let $n_{j\mid a}\ge 0$ denote an activity-specific proxy for destination attractiveness at cell $j$: residential population for \emph{home}, business POIs for \emph{work}, and non-business POIs for \emph{stay-out}. We write $r_{ij}$ for the distance between cells $i$ and $j$. At fine spatial resolution, a standard radiation model can assign an unrealistically large share of an origin's outflow to a single neighboring cell. To reduce this concentration, we define the origin mass as the attractiveness within a characteristic radius $r_0(i,t)$ around $i$:
\begin{equation}
m_{i\mid a}(r_0)=\sum_{k:\,r_{ik}<r_0} n_{k\mid a}.
\label{eq:origin_mass}
\end{equation}
We then aggregate attractiveness by distance shells around $i$ and write $s_{ij\mid a}$ for the cumulative attractiveness strictly closer to $i$ than destination $j$ for activity $a$; the shell definitions are given in Additional file 1, Section S1.1. With $m_{i\mid a}(r_0)$, $s_{ij\mid a}$, and $n_{j\mid a}$, the activity-specific radiation kernel is
\begin{equation}
f_{ij\mid a}(r_0)
=\frac{m_{i\mid a}(r_0)\,n_{j\mid a}}
{\bigl(m_{i\mid a}(r_0)+s_{ij\mid a}\bigr)\bigl(m_{i\mid a}(r_0)+s_{ij\mid a}+n_{j\mid a}\bigr)}.
\label{eq:radiation_kernel}
\end{equation}

The activity-specific kernels do not yet describe changes in destination purpose over the day. We therefore combine them using citywide activity shares $\alpha_{a,t}$ estimated from the Person Trip survey~\cite{PT:online}, with $\sum_a\alpha_{a,t}=1$ for each $t$. This mixture gives an hour-specific score, which we normalize to obtain the destination-choice share for $j\ne i$:
\begin{equation}
f_{ij\mid t}(r_0)=\sum_{a}\alpha_{a,t}\,f_{ij\mid a}(r_0),
\quad
p_{ij\mid t}(r_0)=\frac{f_{ij\mid t}(r_0)}{\sum_{j'\ne i} f_{ij'\mid t}(r_0)}.
\label{eq:mixture_share}
\end{equation}
This explicit normalization makes the shares sum to one over the modeled non-origin destinations without invoking the infinite-population limit of the original radiation model~\cite{simini2012universal}.

The normalized shares determine where trips from an origin are assigned, but not the total amount of flow leaving that origin. We obtain this volume from the origin-specific outflow $T_{i\mid t,c}$. Specifically, we allocate the total data OD outflow for stratum $c$ at time $t$ among origins in proportion to a smoothed origin-side population field $N_{i,t,c}$. The construction of this field and the allocation equation are given in Additional file 1, Section S1.2.

The remaining input is the characteristic radius $r_0(i,t)$, which sets the local distance scale of the destination distribution. We estimate $r_0(i,t)$ once from the pooled OD $T^{\mathrm{data}}_{ij\mid t}=\sum_c T^{\mathrm{data}}_{ij\mid t,c}$ and reuse it for all $c$, whereas quantities such as $T_{i\mid t,c}$ remain stratum-specific. For each candidate radius, we compare the near-field cumulative share in the data OD with the corresponding share from the radiation distribution. We select $r_0(i,t)$ by matching these two shares. Under the approximation detailed in Additional file 1, Section S1.3, this procedure reduces to a median-share condition in the idealized infinite-mass limit. Because the data OD is sparse at 125\,m--hour resolution, the inferred radius can be unstable for weakly supported origin--hour pairs. We therefore smooth the $r_0(i,t)$ field on the grid before constructing the baseline. The cumulative-share equations, approximation, and smoothing bandwidth are given in Additional file 1, Section S1.3. In the Results, the curve labeled $r_0\to0^+$ is a diagnostic based on the original radiation baseline in the positive-zero limit of the characteristic radius.

With the destination-choice share, origin outflow, and characteristic radius in place, the radiation baseline is
\begin{equation}
T^{\mathrm{rad}}_{ij\mid t,c}=T_{i\mid t,c}\,p_{ij\mid t}\!\bigl(r_0(i,t)\bigr).
\label{eq:t_rad}
\end{equation}

Finally, walking and non-walking movements have distinct distance structures, so we construct the radiation baseline separately for the two mode groups. The Tokyo analysis uses the walking/non-walking total adjustment described in Additional file 1, Section~S1.4, whereas the Fukuoka application uses the released mode totals without this adjustment, as detailed in Additional file 1, Section S5.

\subsection{Bayesian update}

The second stage combines $T^{\mathrm{rad}}_{ij\mid t,c}$ and $T^{\mathrm{data}}_{ij\mid t,c}$. Population expansion can amplify sample-level variation in released mobility data, particularly for low-flow entries~\cite{toole2015path,iqbal2014development,fekih2021data}. This motivates a reliability-aware update that balances the released value against the radiation baseline. Fine-grid OD flows can also contain sharp spatial discontinuities arising from local land-use heterogeneity, network constraints, and station-centered access and transfer patterns. We therefore apply the update independently to each OD cell and do not smooth neighboring entries in the OD matrix.

We use a Gamma--Poisson working model in which sampling variation in the data OD is characterized by the effective fluctuation scale $\Delta T_{\mathrm{eff}}$ and the radiation baseline supplies the mode of a Gamma prior~\cite{spiess1987maximum,cascetta1988unified}. The full derivation is given in Additional file 1, Section S2. We call the resulting discrepancy-adaptive estimate the refined estimate:
\begin{equation}
T^{\mathrm{est}}_{ij\mid t,c}
=
w^{\mathrm{data}}_{ij\mid t,c}\,T^{\mathrm{data}}_{ij\mid t,c}
+
\left(1-w^{\mathrm{data}}_{ij\mid t,c}\right)T^{\mathrm{rad}}_{ij\mid t,c}.
\label{eq:t_est}
\end{equation}
The coefficient assigned to the data OD is the data weight,
\begin{equation}
w^{\mathrm{data}}_{ij\mid t,c}
=\frac{\kappa_{ij\mid t,c}}{1+\kappa_{ij\mid t,c}},
\qquad
\kappa_{ij\mid t,c}
=
\frac{\kappa^\ast}{\Delta T_{\mathrm{eff}}}
\left|T^{\mathrm{data}}_{ij\mid t,c}-T^{\mathrm{rad}}_{ij\mid t,c}\right|,
\label{eq:data_weight}
\end{equation}
Here $\kappa_{ij\mid t,c}$ is the mixing ratio, $\kappa^\ast$ is the global dimensionless mixing parameter, and $\Delta T_{\mathrm{eff}}$ is the effective fluctuation scale. The data weight lies between zero and one. A larger absolute discrepancy indicates greater unreliability of the radiation baseline, increases the mixing ratio, and assigns more weight to the data OD, preventing the radiation baseline from overriding a strong empirical departure. When the two inputs coincide, the refined estimate equals their common value.

In the analysis table, masked entries and zero-valued entries are treated identically as $T^{\mathrm{data}}_{ij\mid t,c}=0$, and Eq.~\eqref{eq:t_est} is applied to both. The Tokyo maps use the nominal setting $\Delta T_{\mathrm{eff}}=1$ and $\kappa^\ast=0.945$; the corresponding Fukuoka setting is $\Delta T_{\mathrm{eff}}=1$ and $\kappa^\ast=0.780$. The controlled calibration underlying these settings is described next.

\subsection{Controlled thinning calibration and performance assessment}

We calibrate the Bayesian update under controlled sparsification. For notational simplicity, we suppress the stratum index and denote the unthinned data OD by $T^{\mathrm{data}}_{ij\mid t}$. Within this experiment, $T^{\mathrm{data}}_{ij\mid t}$ plays the role of the latent reference table from which we generate pseudo-observed tables. The experiment therefore measures recovery of this reference under imposed degradation. For each thinning rate $\rho\in[0.05,1]$, we apply Poisson thinning and then rescale by the inverse thinning rate:
\begin{equation}
\widetilde{T}^{(\rho)}_{ij\mid t}
\sim
\mathrm{Poisson}\!\left(\rho T^{\mathrm{data}}_{ij\mid t}\right),
\qquad
T^{(\rho)}_{ij\mid t}
=
\rho^{-1}\widetilde{T}^{(\rho)}_{ij\mid t}.
\label{eq:controlled_thinning}
\end{equation}

The inverse-rate rescaling is a stylized analogue of population renormalization, in which GPS-derived sample counts are expanded to population-scale OD volumes. Under this construction, the expansion preserves the reference mean but increases the variance-to-mean ratio to $\rho^{-1}$. The origin outflows used in the radiation baseline are calculated from the unthinned data OD and are not recalculated from $T^{(\rho)}_{ij\mid t}$. We then apply the radiation--Bayesian update to $T^{(\rho)}_{ij\mid t}$ and compare both $T^{(\rho)}_{ij\mid t}$ and its refined estimate with $T^{\mathrm{data}}_{ij\mid t}$. Although simplified, this experiment captures a central stochastic component of sample-to-population expansion and provides a controlled partial test of whether the refinement attenuates expansion-related error.

For each thinning condition, we generate 10 independent thinned-and-rescaled realizations and estimate the effective fluctuation scale $\Delta T_{\mathrm{eff}}^{(\rho)}$ as detailed in Additional file 1, Section S3.

At each thinning condition, we obtain the optimized global mixing parameter $\kappa^\ast$ by minimizing the MAE of the refined estimate relative to $T^{\mathrm{data}}_{ij\mid t}$. We compare this calibrated refinement with the no-refinement limit $\kappa^\ast=\infty$, which assigns full weight to an independently resampled pseudo-observed table. By comparing the two MAEs, we assess whether the refinement recovers the unthinned data OD more accurately than the pseudo-observed table alone.

\section{Results}
We report results for the Tokyo OD release over the study domain defined above: OD pairs for which at least one endpoint lies within Tokyo's 23 special wards and the other endpoint is within 50\,km of the 23-ward boundary. Additional file 1, Section S5, reports the corresponding sequence of OD maps, data-weight diagnostics, distance profiles, controlled-thinning results, and radiation-baseline diagnostics for Fukuoka Prefecture.

\subsection{Refined OD and data-weight patterns}

We first compare the data OD, radiation baseline, and refined estimate for two representative railway-station areas in Tokyo, Shinagawa and Akabane, at 7:00, 13:00, and 17:00, using $\Delta T_{\mathrm{eff}}=1$ and $\kappa^\ast=0.945$. For each origin, the nine OD maps use a common logarithmic scale; all data-weight maps use a common 0--1 scale.

\begin{figure}[!htbp]
  \centering
  \includegraphics[width=\linewidth]{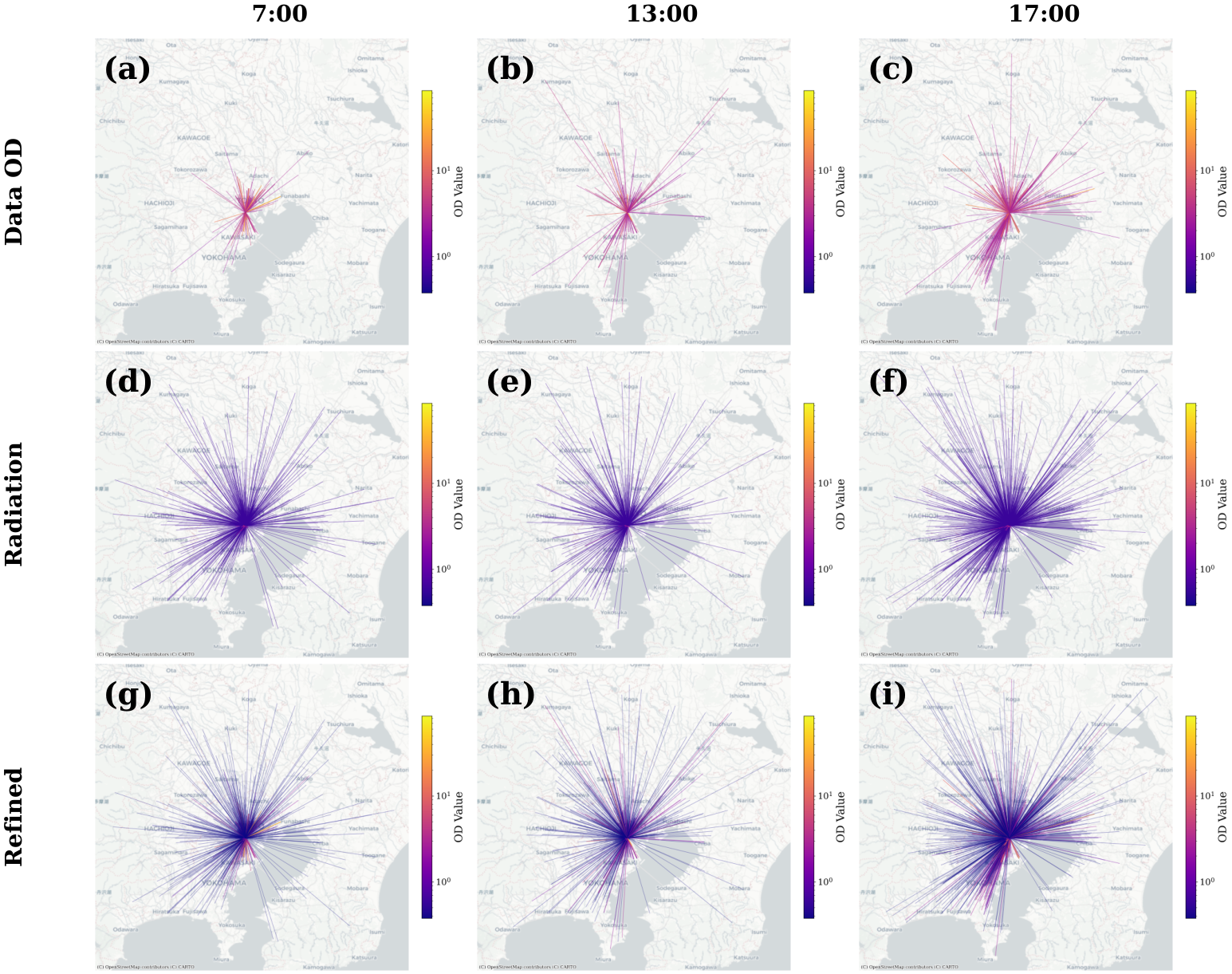}

  \caption{\textbf{Data OD, radiation baseline, and refined estimate for a representative Tokyo railway-station area: Shinagawa.}
  Columns show departure hours 7:00, 13:00, and 17:00; rows show the data OD, radiation baseline, and refined estimate. Link color indicates OD value on a logarithmic scale shared across all nine panels.}
  \label{fig:fig1_od_comparison_shinagawa}
\end{figure}

\begin{figure}[!htbp]
  \centering
  \includegraphics[width=\linewidth]{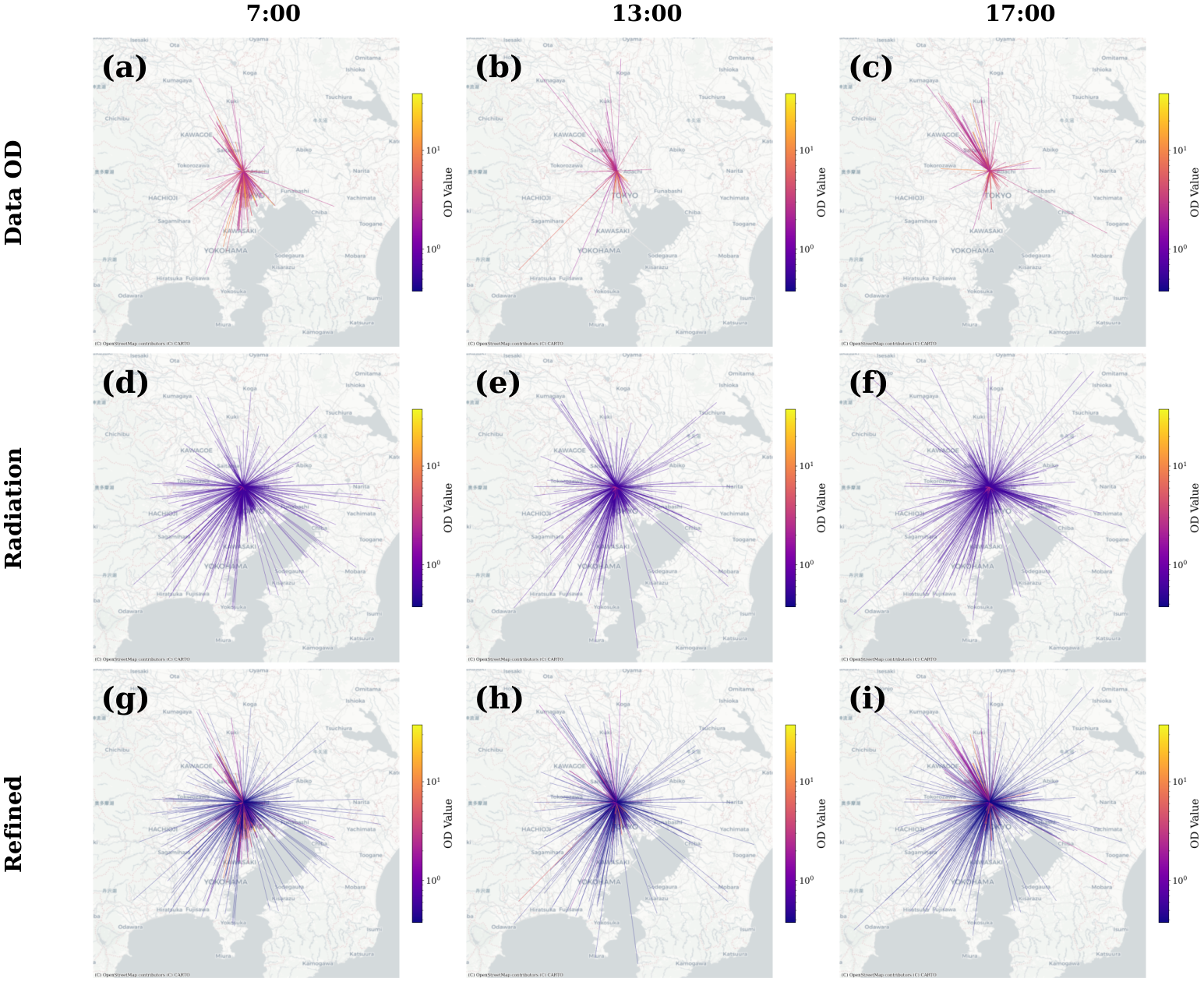}

  \caption{\textbf{Data OD, radiation baseline, and refined estimate for a representative Tokyo railway-station area: Akabane.}
  Columns show departure hours 7:00, 13:00, and 17:00; rows show the data OD, radiation baseline, and refined estimate. Link color indicates OD value on a logarithmic scale shared across all nine panels.}
  \label{fig:fig2_od_comparison_akabane}
\end{figure}

Figs.~\ref{fig:fig1_od_comparison_shinagawa} and \ref{fig:fig2_od_comparison_akabane} show pronounced directional structure in the OD patterns. The data OD contains a comparatively sparse set of observed links, whereas the radiation baseline distributes flow over a broader set of directions and distances. The refined estimate retains high-valued directional features of the data OD while adding lower-intensity, baseline-supported links. Across the three departure hours, the directional emphasis and radial reach vary within each origin, and Shinagawa and Akabane exhibit qualitatively distinct spatial configurations.

\begin{figure}[!htbp]
  \centering
  \includegraphics[width=\linewidth]{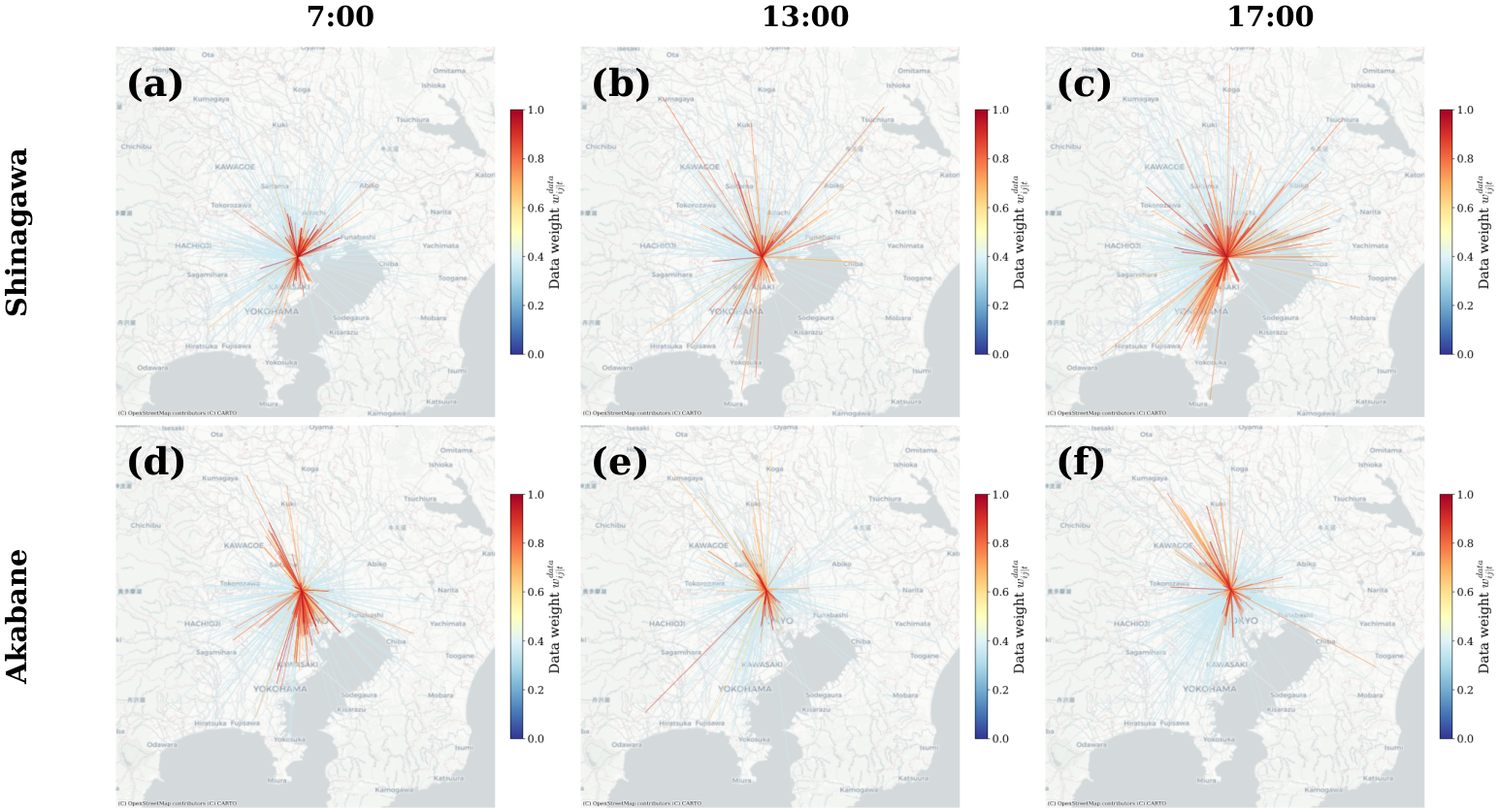}

  \caption{\textbf{Data-weight patterns across representative origins and departure hours.}
  Columns show departure hours 7:00, 13:00, and 17:00; the top row shows Shinagawa and the bottom row shows Akabane. Link color indicates the data weight $w^{\mathrm{data}}_{ij\mid t}$ assigned to the data OD. All six panels use a common scale from 0 to 1.}
  \label{fig:fig3_data_weights}
\end{figure}

Fig.~\ref{fig:fig3_data_weights} maps the relative contributions of the data OD and radiation baseline to the refined patterns. Across all six cases, links with high data weights form directionally concentrated subsets, while low and intermediate data weights occupy a broader radial field. The refined patterns therefore combine a data-anchored backbone along selected directions with a more widely distributed baseline component. The orientation and radial reach of these components vary across departure hours and between origins, showing spatial and temporal variation in the balance between the data OD and radiation baseline.

\begin{figure}[!htbp]
  \centering
  \includegraphics[width=0.78\linewidth]{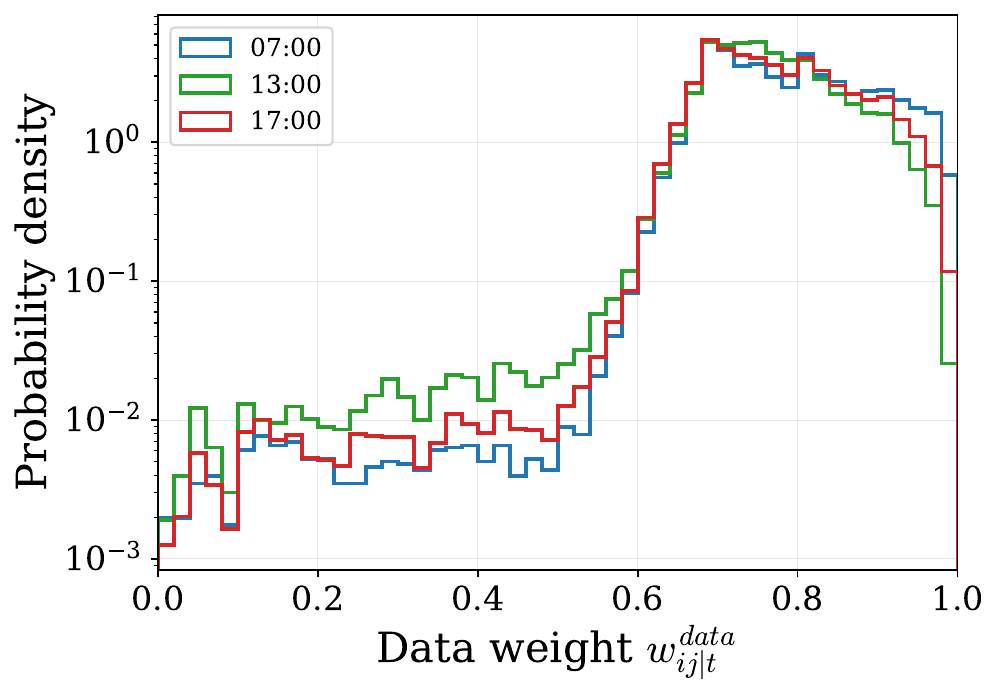}
  \caption{\textbf{Data-weight distributions for positive-flow OD cells by departure hour.}
  For each departure hour, the curve shows the probability density function of $W=w^{\mathrm{data}}_{ij\mid t}$ over cells satisfying $T^{\mathrm{data}}_{ij\mid t}>0$. The horizontal axis is linear, the vertical axis is logarithmic, and all three distributions use the common nominal setting $\Delta T_{\mathrm{eff}}=1$ and $\kappa^\ast=0.945$.}
  \label{fig:fig4_data_weight_pdf}
\end{figure}

Fig.~\ref{fig:fig4_data_weight_pdf} shows that, although most positive-flow OD cells have data weights between 0.6 and 1.0, all three distributions extend below 0.6. These lower-weight cases show that the refinement is not limited to masked or zero-valued entries: the radiation baseline can also make a substantive contribution to a subset of observed positive flows. The distributions further show a clear temporal reversal: above 0.9, the density is highest at 7:00, followed by 17:00 and 13:00, whereas below 0.6 the ordering is reversed. Positive-flow cells are therefore most strongly data-anchored in the morning and, to a lesser extent, in the evening, while the radiation baseline plays a larger relative role at 13:00. This contrast is consistent with the coherent collective flows associated with commuting periods and the more weakly correlated, nearly independent movement directions observed during the afternoon~\cite{shida2020universal}. The adaptive refinement therefore preserves strongly structured commuter flows while allowing a greater model contribution when daytime mobility is less collectively organized.

\subsection{Distance-profile diagnostics}

\begin{figure}[!htbp]
  \centering
  \includegraphics[width=\linewidth]{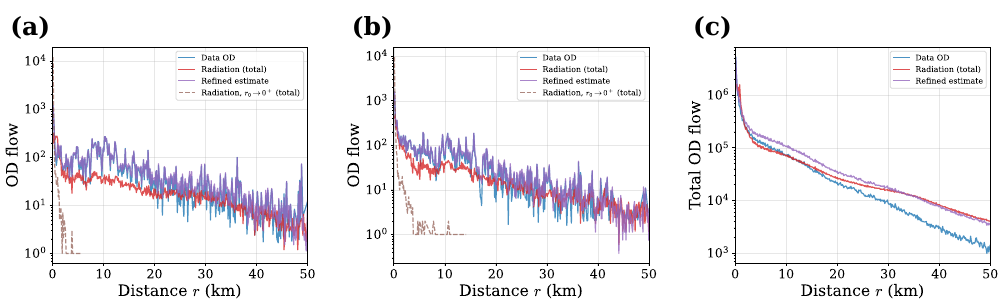}

\caption{\textbf{Distance profiles for representative origins and the global OD distribution.}
\textbf{(a,b)} Full-day profiles for Shinagawa (a) and Akabane (b), comparing the data OD, radiation baseline, $r_0\to0^+$ diagnostic, and refined estimate $T^{\mathrm{est}}$. The $r_0\to0^+$ diagnostic concentrates mass at very short distances, whereas the radiation baseline captures the long-distance tail. \textbf{(c)} Full-day global profiles aggregated over all OD pairs. The refined estimate exceeds both aggregate input curves at intermediate distances and approaches the radiation baseline in the far tail.}
\label{fig:fig5_distance_profiles}
\end{figure}

Fig.~\ref{fig:fig5_distance_profiles}(a,b) shows the distance profiles for Shinagawa and Akabane. At intermediate distances ($r\approx 3$--$20$\,km), the data OD exceeds the radiation baseline by up to a few-fold, whereas agreement becomes strong for $r\gtrsim 20$\,km. In both origin-specific profiles, $T^{\mathrm{est}}$ remains close to the data OD where empirical support is strong, while the baseline complements weakly supported, low-intensity parts of the distribution.

The $r_0\to0^+$ diagnostic represents the positive-zero limit of the characteristic radius and hence the original radiation formulation without the $r_0$-dependent local mass. It concentrates most probability mass at very short distances and fails to reproduce the empirical profile. This comparison shows that a nonzero characteristic scale $r_0(i,t)$ is essential within this formulation for capturing the observed distance dependence, yielding a practically useful baseline especially in the long-distance regime.

At the city scale, Fig.~\ref{fig:fig5_distance_profiles}(c) reports the full-day distance distribution aggregated over all OD pairs. The data OD and the radiation baseline agree closely up to about $20$\,km, beyond which the baseline assigns more mass to long-distance movements. One plausible explanation is distance-dependent thinning in the low-count regime: long-distance flows are dispersed over more OD pairs and are therefore more likely to fall below the suppression threshold $T_{\mathrm{th}}$. The Fukuoka-wide profile likewise shows increasing separation toward long distances (Additional file 1, Fig.~S7(c)), indicating that this pattern is not unique to Tokyo. Additional mechanisms may include network- and time-cost frictions not represented in the baseline.

Although each cell-wise estimate lies between its two inputs, the aggregated distance profile can exceed both input profiles because the data weights vary across OD pairs. Over $5\lesssim r\lesssim 30$\,km, the update preserves large empirical flows in data-dominated pairs while reinstating baseline-supported mass across sparse pairs in the same distance bins, causing $T^{\mathrm{est}}$ to exceed both aggregate input curves. For $r\gtrsim 30$\,km, the distance bins are dominated by low-intensity OD pairs for which the data provide limited constraints, so $T^{\mathrm{est}}$ approaches the radiation baseline.

\subsection{Calibration and performance under controlled thinning}

\begin{figure}[!htbp]
  \centering
  \includegraphics[width=\linewidth]{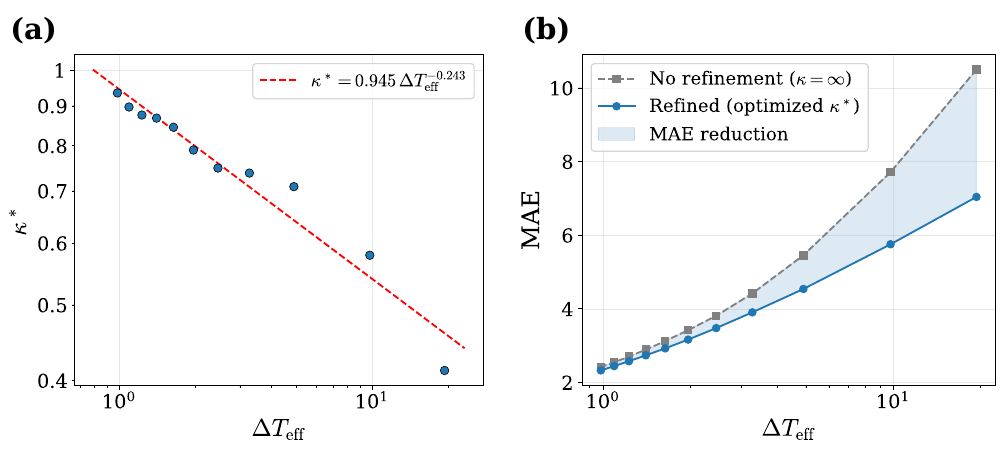}

\caption{\textbf{Calibration and performance of the reliability mechanism under Poisson thinning.}
\textbf{(a)} Condition-specific optimized $\kappa^\ast$ as a function of the effective fluctuation scale $\Delta T_{\mathrm{eff}}$ on log--log axes, with the fitted relation $\kappa^\ast\simeq0.945\,\Delta T_{\mathrm{eff}}^{-0.243}$. \textbf{(b)} Realization-averaged MAE relative to the unthinned data OD for the refined estimate using the condition-specific optimized $\kappa^\ast$ and for an independently resampled no-refinement pseudo-observed table ($\kappa^\ast=\infty$). The shaded band shows the difference between the two curves.}
\label{fig:fig6_controlled_thinning}
\end{figure}

Across thinning conditions, the condition-specific optimized $\kappa^\ast$ decreases with the effective fluctuation scale and follows the shallow power-law relation $\kappa^\ast\simeq0.945\,\Delta T_{\mathrm{eff}}^{-0.243}$ (Fig.~\ref{fig:fig6_controlled_thinning}(a)). This relation gives the nominal value $\kappa^\ast=0.945$ at $\Delta T_{\mathrm{eff}}=1$ used in the Tokyo analyses.

Fig.~\ref{fig:fig6_controlled_thinning}(b) compares realization-averaged MAE relative to the unthinned data OD between the refined estimate using the condition-specific optimized $\kappa^\ast$ and an independently resampled no-refinement pseudo-observed table ($\kappa^\ast=\infty$). The refinement yields lower MAE across the tested range, with the shaded band showing the difference between the two curves. Each pseudo-observed table is expanded back to population scale before refinement; the lower MAE therefore suggests that the framework can attenuate stochastic errors propagated when GPS-sample OD counts are renormalized to a target population, at least under uniform random sampling and known expansion factors. Because $\kappa^\ast$ is optimized against the same unthinned reference, this comparison provides calibration-sample evidence under imposed thinning rather than independent validation or proof of recovery of the true latent OD.

\subsection{Diagnostics of the radiation baseline}

\begin{figure}[!htbp]
  \centering
  \includegraphics[width=\linewidth]{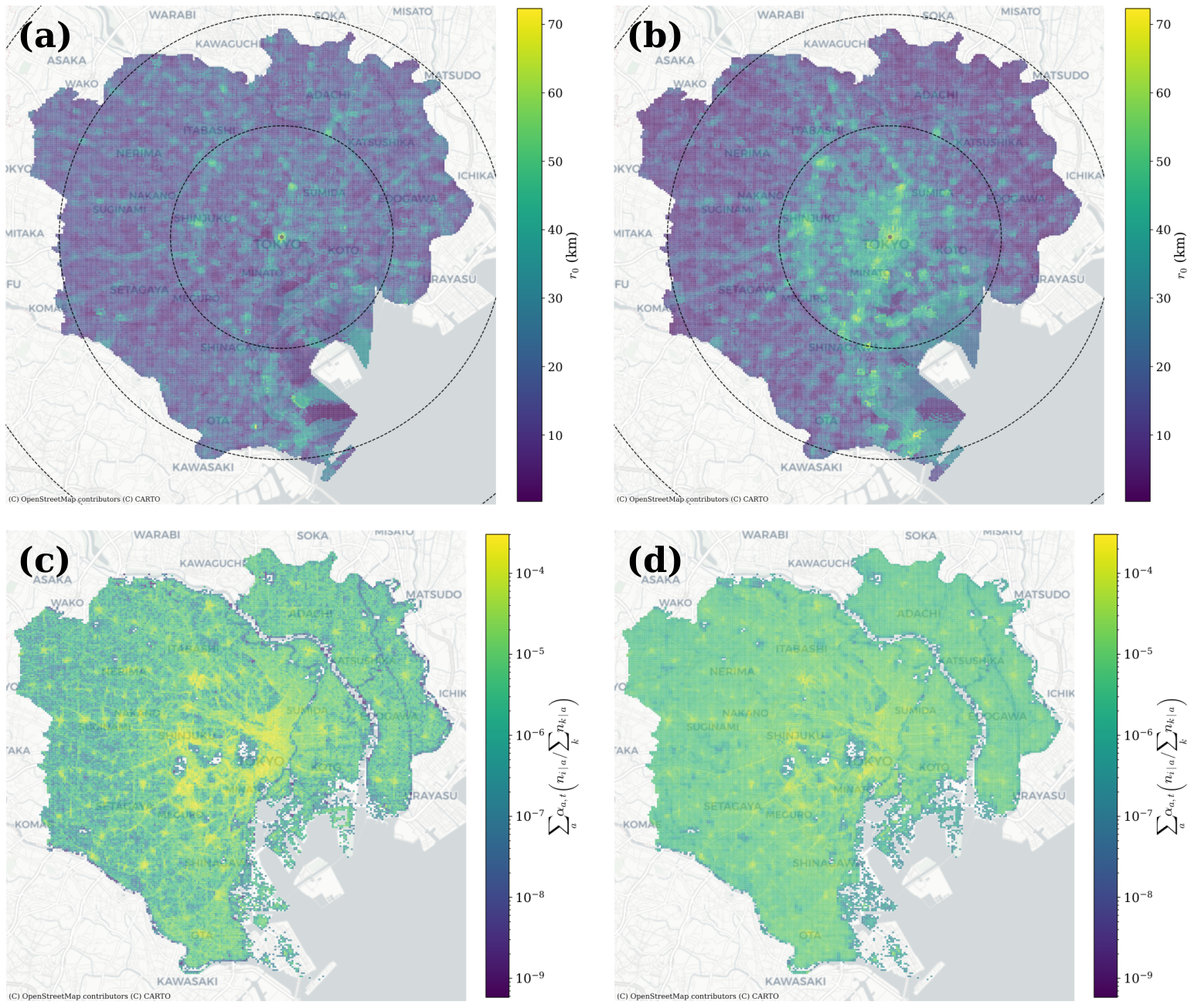}

  \caption{\textbf{Spatial diagnostics of the radiation baseline in Tokyo.}
  \textbf{(a,b)} Smoothed non-walking $r_0(i,t)$ on the 125\,m grid for departures during 7:00--7:59 (a) and 17:00--17:59 (b); brighter colors indicate larger $r_0$. Dashed concentric circles centered on Tokyo Station are shown at 20\,km intervals. The maps show $r_0(i,t)$ only for origins within the 23 wards, while destinations can lie outside this domain. \textbf{(c,d)} Hour-specific weighted attractiveness fields for departures during 7:00--7:59 (c) and 17:00--17:59 (d), given by $\sum_a\alpha_{a,t}\left(n_{i\mid a}/\sum_k n_{k\mid a}\right)$.}
  \label{fig:fig7_prior_diagnostics}
\end{figure}

Fig.~\ref{fig:fig7_prior_diagnostics}(a,b) maps the smoothed non-walking characteristic radius $r_0(i,t)$ for departures during 7:00--7:59 and 17:00--17:59. Elevated values follow major rail corridors and station areas at both times. This alignment is consistent with the role of rail-based access and transfer structures in organizing mobility at broader spatial scales. At 7:00, this structure is concentrated around rail lines and station catchments, whereas at 17:00 larger values extend more broadly across the inner-city core. Because larger $r_0$ denotes a broader characteristic mobility scale, this contrast is consistent with rail-oriented morning mobility and broader evening dispersal. The corresponding distributions before and after Gaussian smoothing are shown in Additional file 1, Fig.~S2.

Fig.~\ref{fig:fig7_prior_diagnostics}(c,d) maps the hour-specific weighted attractiveness field implied by the activity mixture. This field is an input to the radiation baseline rather than OD flow itself. At 7:00, high values concentrate in downtown and around major stations, reflecting the stronger work-related contribution. At 17:00, the emphasis shifts toward residential areas as the home component becomes more prominent.

\section{Discussion and conclusions}

The main contribution of this study is a cell-wise middle ground between accepting a released high-resolution OD table as uniformly informative and replacing it wholesale with a spatial-interaction model. The released table retains sparse, directionally concentrated structures, whereas the radiation baseline distributes flow over a broader set of destinations and distances. Because the update acts independently on each OD pair, the refined estimate can preserve sharp released-flow structure while introducing lower-intensity, baseline-supported links without smoothing neighboring entries. The resulting data weight then makes the relative contribution of the two inputs explicit for every refined entry. This combination of structure preservation and transparent model contribution is the central advantage of the framework.

The temporal data-weight pattern shows that this balance is not constant even within the positive-flow network. Weights are most concentrated near one at 7:00, followed by 17:00, whereas lower weights are more common at 13:00. This ordering mirrors the contrast discussed in the Results between more coherent commuting-period flows and more weakly correlated afternoon movement directions. The data weight is not a calibrated measure of measurement accuracy, so these results do not imply that commuting-period observations are intrinsically more accurate. Rather, they show that the operational balance between released and model-supported structure varies systematically across periods with different collective mobility organization.

The spatial diagnostics reveal a complementary organization across distance and location. For Shinagawa and Akabane, the released OD exceeds the radiation baseline over approximately 3--20 km, while the profiles become more similar beyond approximately 20 km. The \(r_0\to0^+\) diagnostic concentrates flow at very short distances, whereas the inferred \(r_0(i,t)\) broadens the movement range represented by the baseline; larger values also show structured variation around major stations and rail corridors. At the city scale, the baseline increasingly exceeds the released OD at long distances, and a similar separation appears in Fukuoka. Low-count suppression is one plausible contributor because dispersed long-distance flows are more likely to fall below the release threshold, but the present data do not identify a unique cause. Unmodeled network structure, travel-time impedance, and other provider-side processing may also contribute.

The controlled Poisson thinning experiment addresses a different question: whether the refinement behaves as intended when a known stochastic degradation is imposed on the released table. As the effective fluctuation scale increases, the condition-specific optimized \(\kappa^\ast\) decreases, and the calibrated refinement yields lower realization-averaged MAE than the independently resampled no-refinement table throughout the tested range. This systematic response supports the intended weighting behavior under the imposed thinning-and-rescaling process. It does not establish true-flow recovery under real provider sampling or privacy suppression, because the unthinned released OD table serves as the reference and \(\kappa^\ast\) is calibrated against that same reference.

Several limitations define the scope of these results. First, masked entries and ordinary zero-valued entries are encoded identically, so their underlying causes cannot be distinguished. Second, the radiation baseline is data-informed through quantities derived from the released OD product, including origin outflows and the inferred \(r_0(i,t)\), and it also remains conditional on external population and opportunity fields. Third, the update strength depends on the discrepancy between the released table and the baseline, so the Gamma--Poisson formulation should not be interpreted as a clean decomposition between an independent prior and independent observations. Fourth, the data weight is an operational mixing coefficient rather than a direct estimate of measurement reliability. Fifth, the city-specific values of \(\kappa^\ast\) do not establish parameter transferability between Tokyo and Fukuoka. Finally, the cell-wise update does not explicitly represent route choice, travel-time impedance, or network constraints. The refined table should therefore be interpreted as a model-assisted representation of the released OD product rather than an externally validated reconstruction of an inaccessible true OD matrix.

Within these limits, the study provides a practical middle ground between using a released mobility product unchanged and replacing it with a model-based reconstruction. High-resolution OD products can preserve sharply expressed empirical structure while still containing entries, periods, and distance regimes in which the refined representation depends more strongly on structured spatial-interaction support. By making that balance explicit at the cell level, the framework retains the local structure that motivates high-resolution mobility data in the first place while avoiding an all-or-nothing choice between observation and model. This provides a practical route for working with increasingly detailed human-mobility products produced under sampling, population scaling, and privacy constraints.

\backmatter

\bmhead{Supplementary information}
Supplementary Information is included after the main manuscript in this arXiv version. \textbf{Title:} Supplementary Information for Reliability-Aware Refinement of High-Resolution Human Mobility Origin--Destination Tables under Sparsity and Suppression. \textbf{Description:} Extended radiation-baseline definitions and calibration, the Gamma--Poisson derivation, additional Tokyo diagnostics, and the secondary application to Fukuoka Prefecture, including Supplementary Figs.~S1--S10.

\section*{Declarations}

\bmhead{Availability of data and materials}
The aggregated OD dataset (D1) used in this study is proprietary to KDDI Corporation and was used under a commercial data license~\cite{KDDICORP85:online}. The authors are not permitted to redistribute D1. Any request for access to D1 should be directed to KDDI Corporation and is subject to KDDI Corporation's applicable licensing terms. The Person Trip survey (D2), POI data (D3), and population grid (D4) are available from their respective providers under the terms described in~\cite{PT:online,TheFutur61:online,Populati73:online}. The custom analysis software cannot be publicly released because of contractual restrictions associated with the industry collaboration.
\bmhead{Competing interests}
R.S. and T.M. are employees of GEOTRA Co., Ltd. Y.S. and J.O. serve as paid technical advisors to GEOTRA Co., Ltd. under separate agreements concerning general data analysis. These advisory agreements are separate from the collaborative research agreement supporting this study. The authors declare no other competing interests.
\bmhead{Funding}
This work was supported by a collaborative research agreement with GEOTRA Co., Ltd. on improving GEOTRA activity data. GEOTRA Co., Ltd. participated in the research through its employee co-authors, R.S. and T.M.; their individual contributions are described in the Authors' contributions statement.
\bmhead{Authors' contributions}
Yohei Shida: Conceptualization, Methodology, Writing -- original draft, Writing -- review and editing. Ryosuke Susuta: Methodology, Software, Formal analysis, Data curation, Writing -- review and editing. Takuhiro Moriyama: Data curation, Writing -- review and editing. Jun'ichi Ozaki: Conceptualization, Methodology, Writing -- original draft, Writing -- review and editing, Supervision, Project administration.
\bmhead{Acknowledgements}
We thank KDDI Corporation for making the aggregated OD dataset used in this study available.

\bibliography{references}
\end{document}


\title{Supplementary Information for\\Reliability-Aware Refinement of High-Resolution Human Mobility Origin--Destination Tables under Sparsity and Suppression}

\author[1]{\fnm{Yohei} \sur{Shida}}
\equalcont{These authors contributed equally to this work.}
\author[2]{\fnm{Ryosuke} \sur{Susuta}}
\equalcont{These authors contributed equally to this work.}
\author[2]{\fnm{Takuhiro} \sur{Moriyama}}
\author*[3]{\fnm{Jun'ichi} \sur{Ozaki}}\email{ozaki.jun.fe@yokohama-cu.ac.jp}

\affil[1]{\orgdiv{Institute of Systems and Information Engineering}, \orgname{University of Tsukuba}, \orgaddress{\street{1-1-1 Tennodai}, \city{Tsukuba}, \postcode{305-8573}, \state{Ibaraki}, \country{Japan}}}
\affil[2]{\orgname{GEOTRA Co., Ltd.}, \orgaddress{\street{Otemachi One Tower 6F, 1-2-1 Otemachi}, \city{Chiyoda-ku}, \postcode{100-0004}, \state{Tokyo}, \country{Japan}}}
\affil*[3]{\orgdiv{School of Data Science}, \orgname{Yokohama City University}, \orgaddress{\street{22-2 Seto, Kanazawa-ku}, \city{Yokohama}, \postcode{236-0027}, \state{Kanagawa}, \country{Japan}}}

\abstract{This Supplement provides details of the extended radiation baseline and Gamma--Poisson update. Additional diagnostics are reported for Tokyo, together with a secondary application to Fukuoka Prefecture.}

\keywords{supplementary materials, origin--destination table, Bayesian refinement, controlled thinning, human mobility}

\maketitle
\section{Extended radiation baseline details}
\label{sec:supp_prior}

This section collects the definitions and calibration details supporting the extended radiation baseline presented in the Main manuscript. The Bayesian working model is derived separately in Section~\ref{sec:supp_bayesian}.

\subsection{Distance shells and attractiveness fields}
\label{sec:supp_shells}

To incorporate time-of-day variation and purpose heterogeneity, we introduce activity-specific attractiveness fields. Let $a$ index activity types (home, work, and stay-out), and let $n_{j\mid a}\geq 0$ denote an activity-specific proxy for destination attractiveness at cell $j$ (residential population for home, business points of interest for work, and non-business points of interest for stay-out). We write $r_{ij}$ for the distance between cells $i$ and $j$.

For the extended radiation baseline, we first make explicit the treatment of equal-distance ties by aggregating attractiveness over distance shells around each origin. For a fixed origin $i$ and activity $a$, we define
\begin{equation}
n_{r\mid a}^{(i)}=\sum_{k:\,r_{ik}=r}n_{k\mid a},
\qquad
s_{ir\mid a}=\sum_{0<r'<r}n_{r'\mid a}^{(i)},
\qquad
s_{ij\mid a}=s_{i\,r_{ij}\mid a}.
\label{eq:supp_shells}
\end{equation}
To prevent a single neighboring cell from absorbing an unrealistically large share of the outflow, we define the origin mass as the attractiveness contained within a characteristic radius $r_0$ around $i$:
\begin{equation}
m_{i\mid a}(r_0)=\sum_{k:\,r_{ik}<r_0}n_{k\mid a},
\label{eq:supp_origin_mass}
\end{equation}
With $m_{i\mid a}(r_0)$, $s_{ij\mid a}$, and $n_{j\mid a}$, the activity-specific radiation kernel is
\begin{equation}
f_{ij\mid a}(r_0)
=
\frac{m_{i\mid a}(r_0)n_{j\mid a}}
{\bigl(m_{i\mid a}(r_0)+s_{ij\mid a}\bigr)
 \bigl(m_{i\mid a}(r_0)+s_{ij\mid a}+n_{j\mid a}\bigr)}.
\label{eq:supp_radiation_kernel}
\end{equation}

We next introduce time dependence by mixing the activity-specific kernels with citywide activity shares $\alpha_{a,t}$, estimated from the Person Trip survey, satisfying $\sum_a\alpha_{a,t}=1$ for each $t$. We define the mixture score and the normalized destination-choice share as
\begin{equation}
f_{ij\mid t}(r_0)=\sum_a\alpha_{a,t}f_{ij\mid a}(r_0),
\qquad
p_{ij\mid t}(r_0)=\frac{f_{ij\mid t}(r_0)}{\sum_{j'\ne i}f_{ij'\mid t}(r_0)}.
\label{eq:supp_mixture_share}
\end{equation}
This explicit normalization over non-origin destinations avoids relying on the idealized infinite-population argument of the original radiation model.

\subsection{Origin-side outflow construction}
\label{sec:supp_origin_outflow}

We next describe the construction of the origin-side population field used in the Main manuscript. Here $\mathcal{S}[\cdot,\sigma]$ denotes Gaussian smoothing on the grid with bandwidth $\sigma$, and $N^{\mathrm{data}}_{i,t,c}$ denotes the observed presence of stratum $c$ at $(i,t)$. We construct $N_{i,t,c}$ as the product of a smoothed total presence field and a smoothed attribute proportion field. This two-scale construction stabilizes the total volume field at fine resolution while avoiding noisy cell-level demographic fractions in sparsely supported locations. Specifically,
\begin{equation}
N_{i,t,c}
=
\mathcal{S}\!\left[\sum_{c'}N^{\mathrm{data}}_{i,t,c'},75\,\mathrm{m}\right]
\,
\mathcal{S}\!\left[
\frac{N^{\mathrm{data}}_{i,t,c}}{\sum_{c'}N^{\mathrm{data}}_{i,t,c'}},
300\,\mathrm{m}
\right].
\label{eq:supp_origin_weight}
\end{equation}
With this field in hand, the origin-specific outflow is obtained by allocating the total data OD outflow in proportion to $N_{i,t,c}$:
\begin{equation}
T_{\mathrm{total}\mid t,c}=\sum_i\sum_jT^{\mathrm{data}}_{ij\mid t,c},
\qquad
T_{i\mid t,c}
=
T_{\mathrm{total}\mid t,c}\frac{N_{i,t,c}}{\sum_iN_{i,t,c}}.
\label{eq:supp_origin_total}
\end{equation}

\subsection{Characteristic-radius calibration and smoothing}
\label{sec:supp_r0}

We next describe the calibration of $r_0(i,t)$. For each baseline, we estimate $r_0(i,t)$ once from the data OD pooled across attribute strata, $T^{\mathrm{data}}_{ij\mid t}=\sum_c T^{\mathrm{data}}_{ij\mid t,c}$, and reuse it for all $c$. For the following equations, $T^{\mathrm{data}}_{ij\mid t}$ and $T^{\mathrm{rad}}_{ij\mid t}$ denote the pooled data OD and the corresponding radiation baseline used in the calibration. For a candidate radius $r$, we define the near-field fraction of data OD outflow from $i$ at time $t$ and the corresponding fraction under the radiation baseline by
\begin{equation}
R_{\mathrm{data}}(i,t;r)
=
\frac{\sum_{j:\,r_{ij}<r}T^{\mathrm{data}}_{ij\mid t}}
{\sum_jT^{\mathrm{data}}_{ij\mid t}},
\qquad
R_{\mathrm{rad}}(i,t;r)
=
\frac{\sum_{j:\,r_{ij}<r}T^{\mathrm{rad}}_{ij\mid t}}
{\sum_jT^{\mathrm{rad}}_{ij\mid t}}.
\label{eq:supp_r_cumulative}
\end{equation}
We then impose the calibration condition
\begin{equation}
R_{\mathrm{data}}(i,t;r_0)=R_{\mathrm{rad}}(i,t;r_0).
\label{eq:supp_r0_condition}
\end{equation}
To obtain a computable approximation, we evaluate $R_{\mathrm{rad}}(i,t;r)$ at $r=r_0$ under $n_{i\mid a}\ll m_{i\mid a}(r_0)$ and $n_{i\mid a}\ll M_a=\sum_k n_{k\mid a}$, giving
\begin{equation}
R_{\mathrm{data}}(i,t;r_0)
=
\frac{\sum_a\alpha_{a,t}
\dfrac{m_{i\mid a}(r_0)-n_{i\mid a}}
{2m_{i\mid a}(r_0)-n_{i\mid a}}}
{\sum_a\dfrac{\alpha_{a,t}}
{1+m_{i\mid a}(r_0)/(M_a-n_{i\mid a})}}
\simeq
\frac{1}{2}
\left(\sum_a\frac{\alpha_{a,t}}
{1+m_{i\mid a}(r_0)/M_a}\right)^{-1}.
\label{eq:supp_r0_approx}
\end{equation}
Using the calibration condition, this further reduces in the idealized infinite-mass limit $M_a\to\infty$ (i.e., an effectively unbounded attractiveness field) to
\begin{equation}
R_{\mathrm{rad}}(i,t;r_0)\simeq\frac{1}{2}.
\label{eq:supp_r0_half}
\end{equation}
Thus, the calibration reduces to a median-share condition in this limit. Because the data OD is sparse at 125\,m--hour resolution, the inferred $r_0(i,t)$ can be unstable for weakly supported origin--hour pairs. We therefore smooth the inferred $r_0(i,t)$ on the grid with bandwidth $\sigma=300\,\mathrm{m}$ before constructing the baseline used in subsequent analyses.

\subsection{Tokyo mode-separated baseline and walking adjustment}
\label{sec:supp_modes}

For the Tokyo analysis, we use the released mode-disaggregated OD tables to construct the extended radiation baseline separately for walking and non-walking flows so that each mode group has its own characteristic distance scale. The overall radiation baseline $T^{\mathrm{rad}}_{ij\mid t,c}$ is the sum of these two baselines and is used as the input to the Bayesian update.

The mode-disaggregated tables suggested that the walking-mode share was systematically underestimated. We therefore apply a simple origin-wise adjustment to the walking/non-walking split used in the baseline construction. Let
\begin{equation}
W_{i\mid t,c}=T^{(\mathrm{walk})}_{i\mid t,c},
\qquad
V_{i\mid t,c}=T^{(\mathrm{nonwalk})}_{i\mid t,c},
\qquad
T_{i\mid t,c}=W_{i\mid t,c}+V_{i\mid t,c}.
\end{equation}
denote, respectively, the walking and non-walking outflows, and their total, from origin $i$ at departure hour $t$ in stratum $c$. We inflate the walking total by a global factor $a>1$ while conserving the total outflow:
\begin{equation}
W'_{i\mid t,c}=aW_{i\mid t,c},
\qquad
V'_{i\mid t,c}=T_{i\mid t,c}-W'_{i\mid t,c}.
\label{eq:supp_walking_adjustment}
\end{equation}
For each origin, the walking and non-walking baseline rows are rescaled to have totals $W'_{i\mid t,c}$ and $V'_{i\mid t,c}$, respectively, while retaining their within-mode destination shares. The resulting rows are used in the Tokyo baseline analysis.

\begin{figure}[!htbp]
  \centering
  \includegraphics[width=\linewidth]{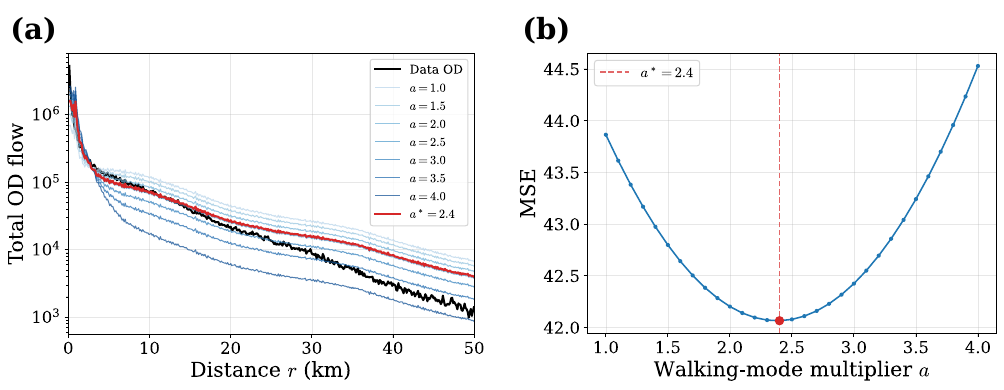}
  \caption{\textbf{Calibration of the walking-mode total adjustment.} Panel (a) shows global distance distributions obtained under different values of the walking inflation factor $a$; the horizontal axis shows distance and the vertical axis shows total OD flow. Panel (b) shows the mean squared error (MSE) between the adjusted radiation baselines and the corresponding data OD tables as a function of $a$. The minimum occurs at $a=2.4$, which is used in the Tokyo analysis.}
  \label{fig:sfig1_walking_adjustment}
\end{figure}

We choose $a$ by comparing the adjusted radiation baselines with the corresponding data OD tables. Figure~\ref{fig:sfig1_walking_adjustment}(a) shows the resulting global distance distributions for different values of $a$, and Fig.~\ref{fig:sfig1_walking_adjustment}(b) reports the corresponding mean squared error (MSE). The minimum occurs at $a=2.4$, which is used throughout the Tokyo analysis when applying mode-separated baselines.

\section{Bayesian working model and MAP derivation}
\label{sec:supp_bayesian}

Because OD flows at 125\,m resolution can exhibit sharp local discontinuities due to local land-use heterogeneity, network constraints, and station-centered access and transfer patterns, the second stage operates independently on each OD cell and does not smooth neighboring entries. We represent sampling variation in the data OD with a Poisson working likelihood governed by the effective fluctuation scale $\Delta T_{\mathrm{eff}}>0$:
\begin{equation}
\frac{T^{\mathrm{data}}_{ij\mid t,c}}{\Delta T_{\mathrm{eff}}}
\sim
\mathrm{Poisson}\!\left(
\frac{T_{ij\mid t,c}}{\Delta T_{\mathrm{eff}}}
\right),
\label{eq:supp_poisson_model}
\end{equation}
where $T_{ij\mid t,c}$ denotes the latent OD flow. Under this working likelihood,
\begin{equation}
\mathbb{E}\!\left[T^{\mathrm{data}}_{ij\mid t,c}\mid T_{ij\mid t,c}\right]
=T_{ij\mid t,c},
\qquad
\mathrm{Var}\!\left[T^{\mathrm{data}}_{ij\mid t,c}\mid T_{ij\mid t,c}\right]
=\Delta T_{\mathrm{eff}}T_{ij\mid t,c}.
\label{eq:supp_delta_teff}
\end{equation}
Thus, $\Delta T_{\mathrm{eff}}$ is the conditional variance-to-mean ratio and represents the fluctuation scale after gross-up and other release processing. Its empirical estimator for the repeated-thinning calibration is given in Section~\ref{sec:supp_thinning}.

For a fixed positive scale $\theta_{ij\mid t,c}$, we encode the radiation baseline $T^{\mathrm{rad}}_{ij\mid t,c}$ as prior information through
\begin{equation}
T_{ij\mid t,c}
\sim
\mathrm{Gamma}\!\left(
1+\frac{T^{\mathrm{rad}}_{ij\mid t,c}}{\theta_{ij\mid t,c}},
\theta_{ij\mid t,c}
\right),
\label{eq:supp_gamma_prior}
\end{equation}
whose mode is $T^{\mathrm{rad}}_{ij\mid t,c}$. In this shape--scale parameterization,
\begin{equation}
\frac{\mathrm{Var}(T_{ij\mid t,c})}
{\mathbb{E}(T_{ij\mid t,c})}
=
\theta_{ij\mid t,c}.
\label{eq:supp_theta_dispersion}
\end{equation}
$\theta_{ij\mid t,c}$ therefore has the same units as the flow.

For each data OD value in the analysis table, Gamma--Poisson conjugacy gives a Gamma posterior with shape and rate
\[
1+\frac{T^{\mathrm{rad}}_{ij\mid t,c}}{\theta_{ij\mid t,c}}
+\frac{T^{\mathrm{data}}_{ij\mid t,c}}{\Delta T_{\mathrm{eff}}},
\qquad
\frac{1}{\theta_{ij\mid t,c}}+\frac{1}{\Delta T_{\mathrm{eff}}},
\]
respectively. Its mode is the refined estimate. Defining the mixing ratio and data weight by
\begin{equation}
\kappa_{ij\mid t,c}
=
\frac{\theta_{ij\mid t,c}}{\Delta T_{\mathrm{eff}}},
\qquad
w^{\mathrm{data}}_{ij\mid t,c}
=
\frac{\kappa_{ij\mid t,c}}{1+\kappa_{ij\mid t,c}}.
\label{eq:supp_kappa_theta}
\end{equation}
gives
\begin{equation}
T^{\mathrm{est}}_{ij\mid t,c}
=
w^{\mathrm{data}}_{ij\mid t,c}T^{\mathrm{data}}_{ij\mid t,c}
+
\left(1-w^{\mathrm{data}}_{ij\mid t,c}\right)
T^{\mathrm{rad}}_{ij\mid t,c}.
\label{eq:supp_map_blend}
\end{equation}
To obtain the discrepancy-adaptive update used in the analysis, we interpret $\theta_{ij\mid t,c}$ as the unreliability scale of the radiation baseline and relate it to the absolute difference between the data OD and radiation baseline:
\begin{equation}
\theta_{ij\mid t,c}
=
\kappa^\ast
\left|
T^{\mathrm{data}}_{ij\mid t,c}
-
T^{\mathrm{rad}}_{ij\mid t,c}
\right|,
\label{eq:supp_theta_heuristic}
\end{equation}
where $\kappa^\ast>0$ is the global dimensionless mixing parameter shared across cells. Because $\theta_{ij\mid t,c}$ is determined from the observed discrepancy, the resulting estimator is a plug-in working rule rather than an update with a data-independent prior scale. Substituting Eq.~\eqref{eq:supp_theta_heuristic} into Eq.~\eqref{eq:supp_kappa_theta} yields
\begin{equation}
\kappa_{ij\mid t,c}
=
\frac{\kappa^\ast}{\Delta T_{\mathrm{eff}}}
\left|
T^{\mathrm{data}}_{ij\mid t,c}
-
T^{\mathrm{rad}}_{ij\mid t,c}
\right|.
\label{eq:supp_kappa_def}
\end{equation}
A larger absolute discrepancy indicates greater unreliability of the radiation baseline, increases the mixing ratio, and assigns more weight to the data OD. If the discrepancy is exactly zero, the Gamma scale in Eq.~\eqref{eq:supp_theta_heuristic} vanishes; the refined estimate is then defined by the continuous limit of Eq.~\eqref{eq:supp_map_blend}, which equals the common data OD and baseline value.

In the analysis table, masked entries and zero-valued entries are treated identically as $T^{\mathrm{data}}_{ij\mid t,c}=0$, and Eq.~\eqref{eq:supp_map_blend} is applied to both.

The Tokyo maps and data-weight distributions use
\begin{equation}
\Delta T_{\mathrm{eff}}=1,
\qquad
\kappa^\ast=0.945.
\label{eq:supp_nominal_setting}
\end{equation}

\section{Estimation of the effective fluctuation scale under controlled thinning}
\label{sec:supp_thinning}

Using the Poisson thinning and inverse-rate rescaling described in the Main manuscript, for each thinning rate $\rho$ we generate $L$ independent realizations $T^{(\rho,\ell)}_{ij\mid t}$. The empirical mean and variance across realizations at each fixed OD cell provide the effective fluctuation scale
\begin{equation}
\Delta T_{\mathrm{eff}}^{(\rho)}
=
\frac{
\left\langle \mathrm{Var}_{\ell}\!\left[T^{(\rho,\ell)}_{ij\mid t}\right]\right\rangle_{i,j,t}
}{
\left\langle \mathbb{E}_{\ell}\!\left[T^{(\rho,\ell)}_{ij\mid t}\right]\right\rangle_{i,j,t}
},
\qquad \ell=1,\ldots,L,\quad L=10.
\label{eq:supp_thinning_fluctuation}
\end{equation}
Here, $\mathbb{E}_{\ell}[\cdot]$ and $\mathrm{Var}_{\ell}[\cdot]$ are taken across the $L$ realizations at fixed $(i,j,t)$, and $\langle X_{ij\mid t}\rangle_{i,j,t}$ averages over the OD cells and hours included in the experiment. Thus, $\Delta T_{\mathrm{eff}}^{(\rho)}$ is the ratio of the average cell-wise variance to the average cell-wise mean, rather than an average of cell-wise variance-to-mean ratios.

\section{Additional Tokyo diagnostics}
\label{sec:supp_tokyo_diagnostics}

\subsection{Characteristic-radius distributions before and after smoothing}
\label{sec:supp_tokyo_r0_distributions}

\begin{figure}[!htbp]
  \centering
  \includegraphics[width=\linewidth]{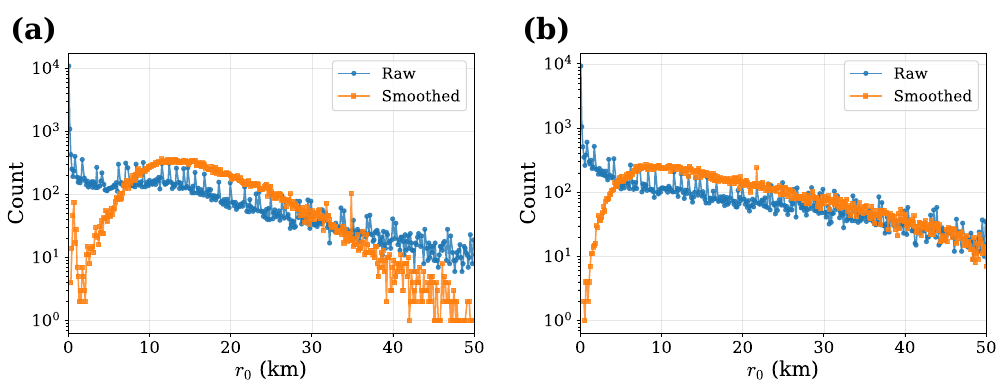}

  \caption{\textbf{Characteristic-radius distributions before and after Gaussian smoothing in Tokyo.} Panels show the distributions of $r_0$ at 7:00 (a) and 17:00 (b), before (raw) and after Gaussian smoothing with a $300\,\mathrm{m}$ bandwidth. The vertical axis is the binned count of origin cells on a logarithmic scale and the horizontal axis is $r_0$ (km). Gaussian smoothing compresses the distribution by shrinking extreme values toward the center, reducing the overall dispersion of $r_0$ and damping noise-driven spikes from weakly supported origin--hour pairs.}
  \label{fig:sfig2_r0_distributions}
\end{figure}

Figure~\ref{fig:sfig2_r0_distributions}(a,b) compares the empirical distributions of the inferred non-walking characteristic radius $r_0$ at 7:00 and 17:00 before and after Gaussian smoothing. At both times, smoothing reduces the overall dispersion and damps noise-driven spikes associated with weakly supported origin--hour pairs, stabilizing the mesoscale $r_0$ field used in the Tokyo analysis.

\section{Secondary application to Fukuoka Prefecture}
\label{sec:supp_fukuoka}

We report a companion analysis for Fukuoka Prefecture using the same sequence of analyses as for Tokyo: a mode-separated extended radiation baseline followed by cell-wise Bayesian refinement, hourly OD and data-weight diagnostics, distance profiles, controlled thinning, and spatial diagnostics of the baseline. Unless stated otherwise, definitions, notation, and estimation details are the same as in the preceding sections. The observation period covers the 22 weekdays in June 2023, and the resulting OD tables are normalized by the number of weekdays. OD pairs are included if either endpoint lies within the target area and the other is within $20\,\mathrm{km}$ of its boundary. The displayed estimates use $\Delta T_{\mathrm{eff}}=1$ and $\kappa^\ast=0.780$.

As in Tokyo, we construct the Fukuoka radiation baseline separately for walking and non-walking flows and sum the two components before Bayesian refinement. Unlike Tokyo, Fukuoka uses the released walking and non-walking totals without the walking-mode total adjustment in Eq.~\eqref{eq:supp_walking_adjustment}; the factor $a=2.4$ is not applied.

\begin{figure}[!htbp]
  \centering
  \includegraphics[width=\linewidth]{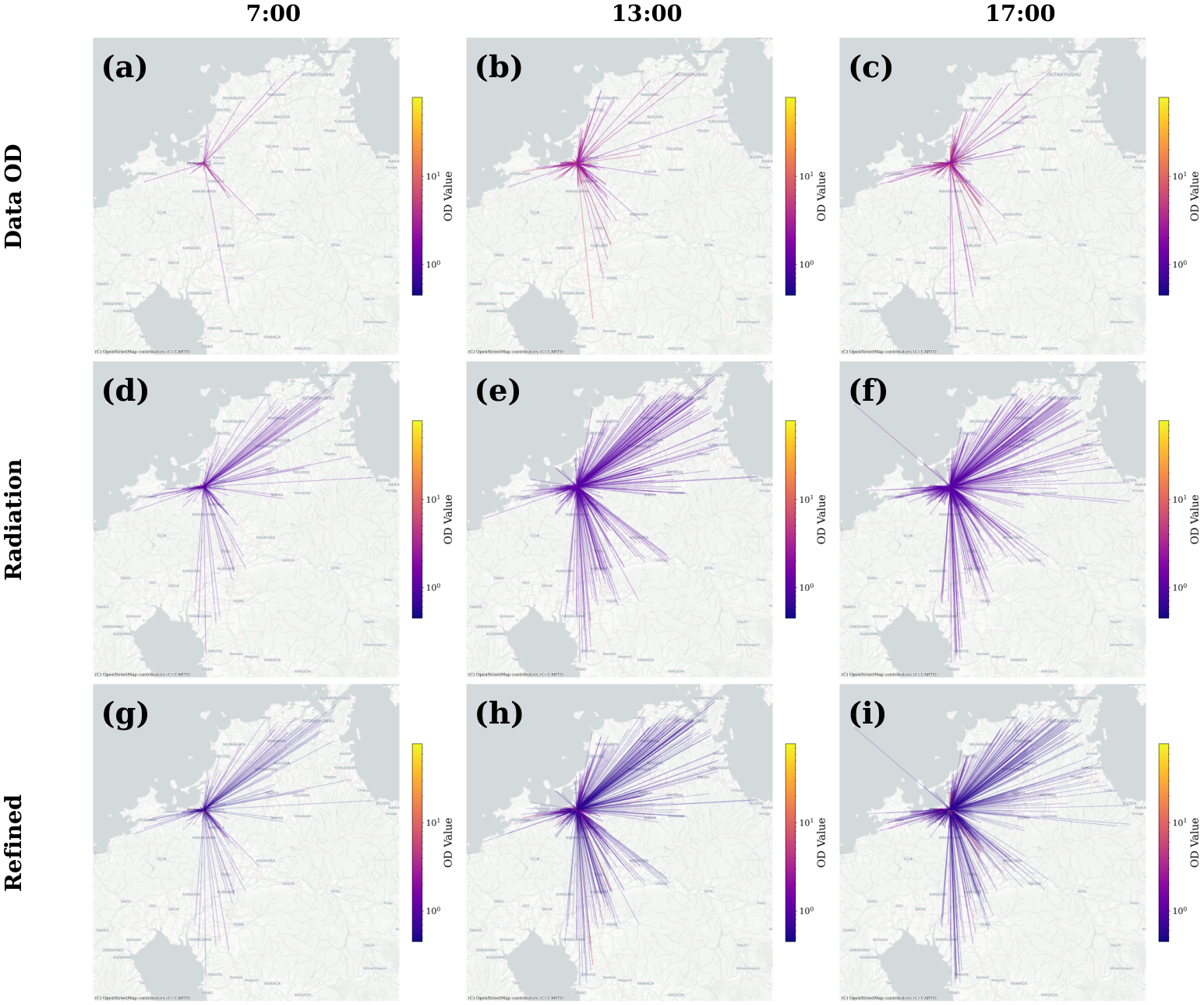}
  \caption{\textbf{Data OD, radiation baseline, and refined estimate for Hakata in Fukuoka Prefecture.} Columns show departure hours 7:00, 13:00, and 17:00; rows show the data OD, radiation baseline, and refined estimate. Link color indicates OD value on a logarithmic scale shared across all nine panels.}
  \label{fig:sfig3_fukuoka_hakata_od}
\end{figure}

\begin{figure}[!htbp]
  \centering
  \includegraphics[width=\linewidth]{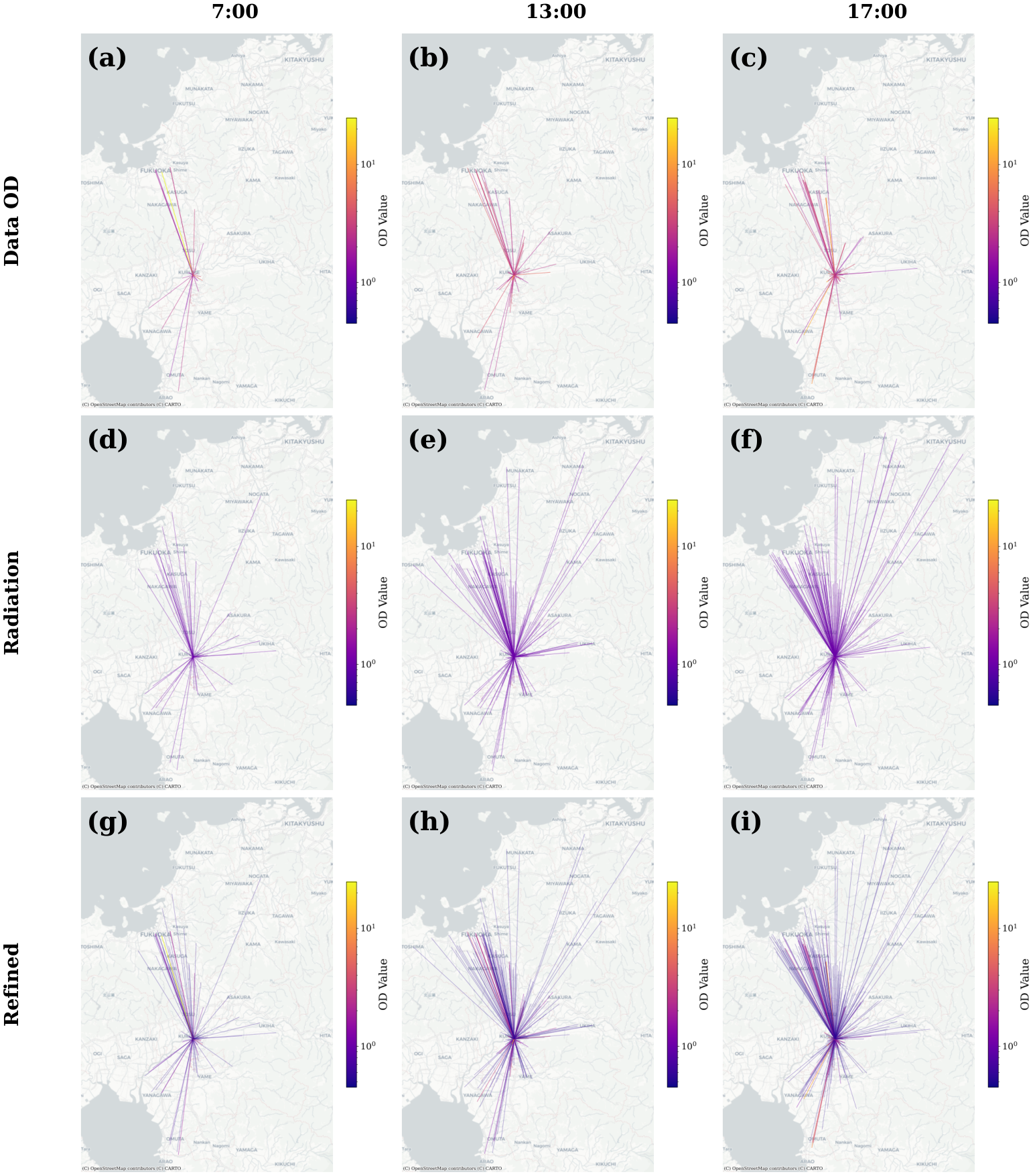}
  \caption{\textbf{Data OD, radiation baseline, and refined estimate for Kurume in Fukuoka Prefecture.} Columns show departure hours 7:00, 13:00, and 17:00; rows show the data OD, radiation baseline, and refined estimate. Link color indicates OD value on a logarithmic scale shared across all nine panels.}
  \label{fig:sfig4_fukuoka_kurume_od}
\end{figure}

Figures~\ref{fig:sfig3_fukuoka_hakata_od} and \ref{fig:sfig4_fukuoka_kurume_od}, corresponding to Main Figs.~1 and 2, compare the data OD, radiation baseline, and refined estimate for Hakata and Kurume at 7:00, 13:00, and 17:00. The radiation baseline supplies a complete radial field, while the refined estimate retains prominent directionally concentrated links in the data OD and introduces baseline-supported flow where the data provide weaker support. The balance varies by origin and departure hour, demonstrating that the cell-wise update remains spatially and temporally adaptive in the Fukuoka setting.

\begin{figure}[!htbp]
  \centering
  \includegraphics[width=\linewidth]{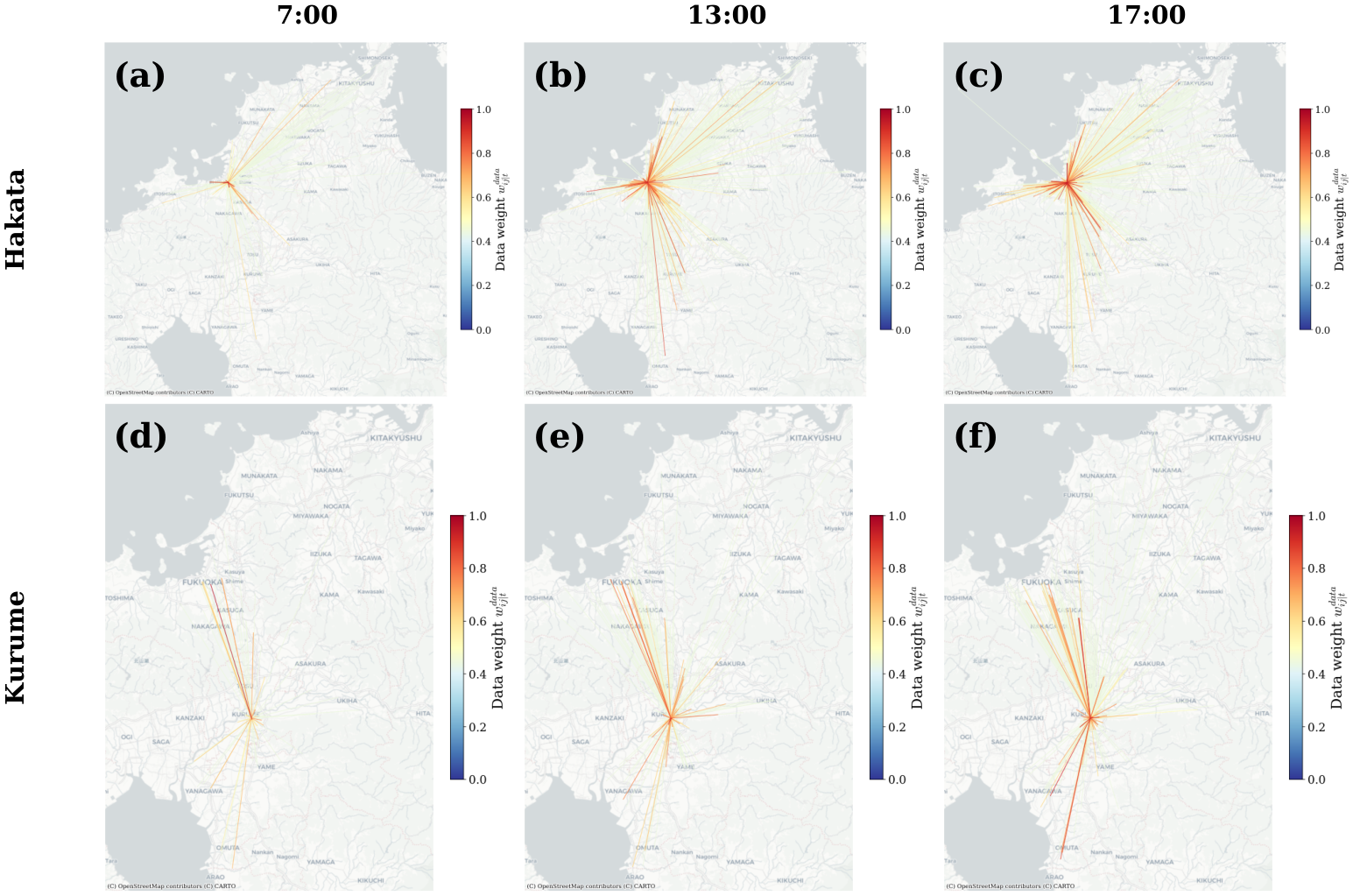}
  \caption{\textbf{Data-weight patterns across representative Fukuoka origins and departure hours.} Columns show departure hours 7:00, 13:00, and 17:00; the top row shows Hakata and the bottom row shows Kurume. Link color indicates the data weight $w^{\mathrm{data}}_{ij\mid t}$ on a common scale from 0 to 1.}
  \label{fig:sfig5_fukuoka_data_weights}
\end{figure}

\begin{figure}[!htbp]
  \centering
  \includegraphics[width=0.78\linewidth]{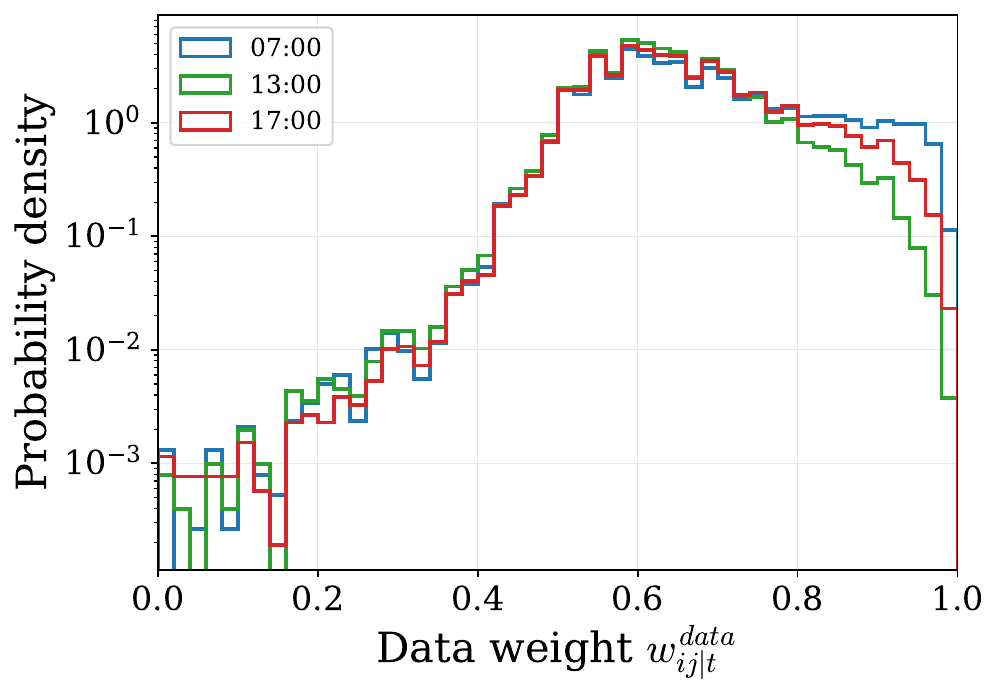}
  \caption{\textbf{Data-weight distributions for positive-flow OD cells in Fukuoka by departure hour.} For each departure hour, the curve shows the probability density function of $W=w^{\mathrm{data}}_{ij\mid t}$ over cells satisfying $T^{\mathrm{data}}_{ij\mid t}>0$. The horizontal axis is linear, the vertical axis is logarithmic, and all three distributions use $\Delta T_{\mathrm{eff}}=1$ and $\kappa^\ast=0.780$.}
  \label{fig:sfig6_fukuoka_data_weight_pdf}
\end{figure}

Figure~\ref{fig:sfig5_fukuoka_data_weights}, corresponding to Main Fig.~3, maps the data weights for the same origins and departure hours. Directionally concentrated links with relatively high weights coexist with a broader field of lower and intermediate weights. Figure~\ref{fig:sfig6_fukuoka_data_weight_pdf}, corresponding to Main Fig.~4, shows the associated distributions over positive-flow OD cells. Most of the density lies at intermediate to high data weights, while all three departure hours retain a lower-weight tail. The 7:00 distribution has the strongest upper tail near one, indicating that positive morning flows are more strongly data-anchored than the corresponding daytime flows. Compared with Tokyo, the Fukuoka distributions are shifted toward lower data weights, indicating that, under the respective nominal calibrations, a larger share of positive-flow OD cells has weaker data support and receives a correspondingly larger contribution from the radiation baseline.

\begin{figure}[!htbp]
  \centering
  \includegraphics[width=\linewidth]{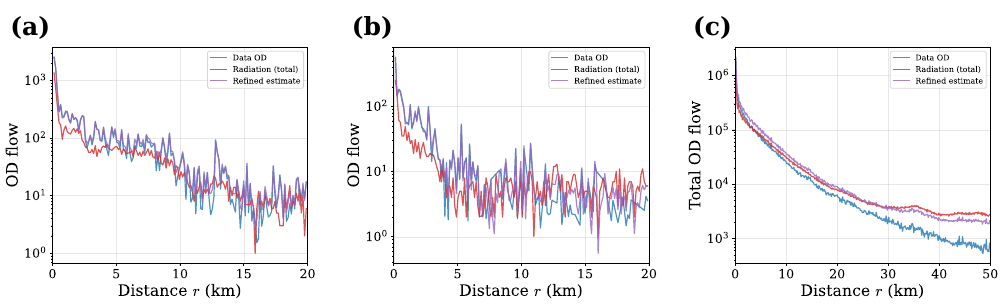}
  \caption{\textbf{Distance profiles for representative origins and the global OD distribution in Fukuoka Prefecture.} Panels compare the data OD, radiation baseline, and refined estimate $T^{\mathrm{est}}$ for Hakata (a), Kurume (b), and the Fukuoka-wide aggregate (c). Panels (a,b) show full-day origin-specific profiles. Panel (c) aggregates all OD pairs over the full day and shows increasing separation between the data OD and radiation baseline toward long distances.}
  \label{fig:sfig7_fukuoka_distance_profiles}
\end{figure}

Figure~\ref{fig:sfig7_fukuoka_distance_profiles}, corresponding to Main Fig.~5, compares the full-day distance profiles for Hakata, Kurume, and the Fukuoka-wide aggregate. For both origins, the radiation baseline is smoother than the data OD and underestimates several pronounced short- and intermediate-distance empirical peaks, while the refined estimate remains close to the data-supported profile. At the Fukuoka-wide scale, the radiation baseline assigns progressively more mass than the data OD toward moderate and long distances (Fig.~\ref{fig:sfig7_fukuoka_distance_profiles}(c)). The refined aggregate retains the short-distance concentration of the data while reinstating baseline-supported mass in weakly supported OD pairs.

\begin{figure}[!htbp]
  \centering
  \includegraphics[width=\linewidth]{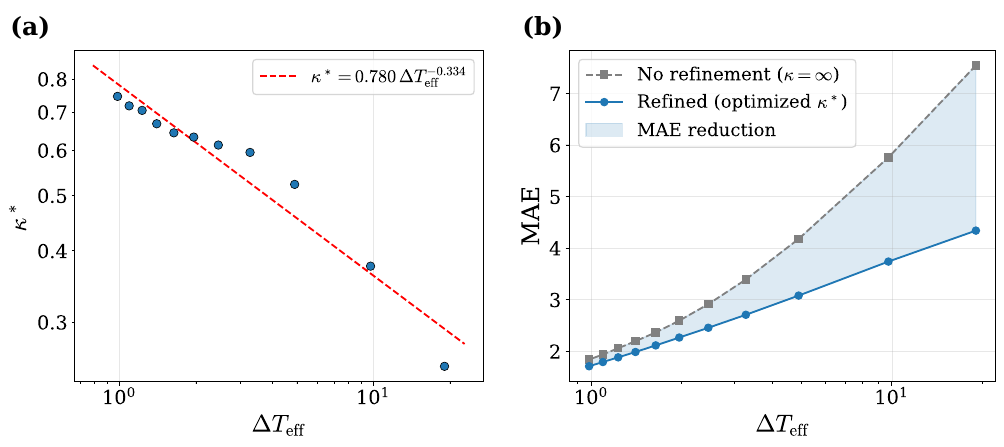}
  \caption{\textbf{Calibration and performance of the reliability mechanism for Fukuoka Prefecture under Poisson thinning.} Panel (a) shows the condition-specific optimized $\kappa^\ast$ as a function of $\Delta T_{\mathrm{eff}}$ on log--log axes, with the fitted relation $\kappa^\ast\simeq0.780(\Delta T_{\mathrm{eff}})^{-0.334}$. Panel (b) shows realization-averaged MAE relative to the unthinned data OD for the refined estimate using the condition-specific optimized $\kappa^\ast$ and for an independently resampled no-refinement pseudo-observed table ($\kappa^\ast=\infty$). The shaded band shows the difference between the two curves.}
  \label{fig:sfig8_fukuoka_thinning}
\end{figure}

We repeated the controlled thinning experiment for Fukuoka Prefecture using the same procedure as for Tokyo: Poisson thinning followed by inverse-rate rescaling, with the unthinned data OD as the reference. This analysis, shown in Fig.~\ref{fig:sfig8_fukuoka_thinning}, corresponds to Main Fig.~6. For each thinning condition, 10 independent thinned-and-rescaled realizations are used to estimate $\Delta T_{\mathrm{eff}}$ and obtain the condition-specific optimized $\kappa^\ast$. The optimized values follow the power-law relation
\begin{equation}
\kappa^\ast
\simeq
0.780\left(\Delta T_{\mathrm{eff}}\right)^{-0.334}.
\label{eq:supp_kappa_diagnostic_fukuoka}
\end{equation}
At $\Delta T_{\mathrm{eff}}=1$, this fitted relation gives the nominal value $\kappa^\ast=0.780$ used in the Fukuoka maps and distributions. Figure~\ref{fig:sfig8_fukuoka_thinning}(b) compares realization-averaged MAE relative to the unthinned data OD between the refined estimate using the condition-specific optimized $\kappa^\ast$ and an independently resampled no-refinement pseudo-observed table ($\kappa^\ast=\infty$). The refinement yields lower MAE across the tested range, with the shaded band showing the difference between the two curves. Because $\kappa^\ast$ is optimized against the same unthinned reference, this comparison provides calibration-sample evidence under imposed thinning rather than independent validation or proof of recovery of the true latent OD.

\begin{figure}[!htbp]
  \centering
  \includegraphics[width=\linewidth]{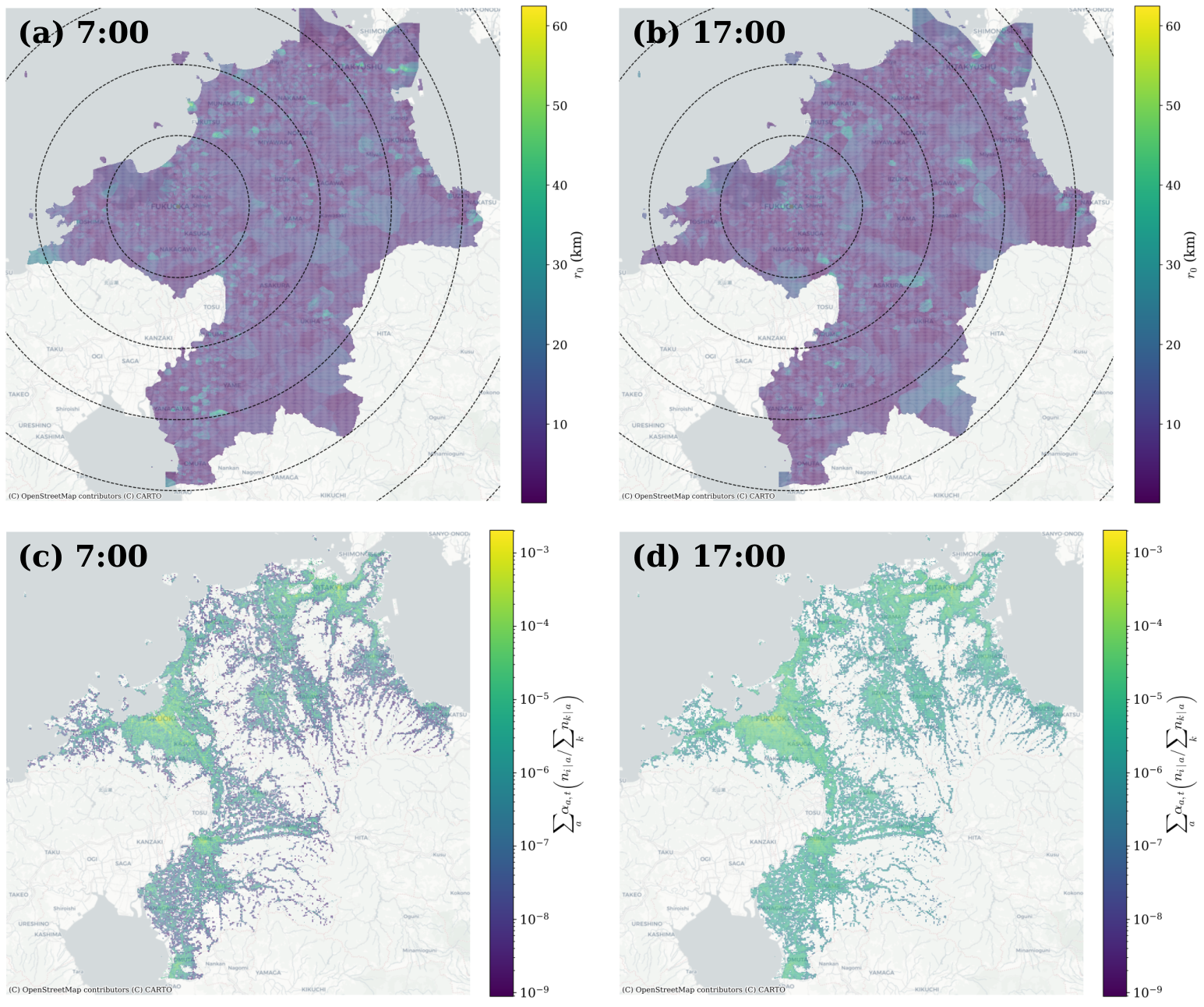}
  \caption{\textbf{Characteristic-radius and weighted-attractiveness fields in Fukuoka Prefecture.} Panels (a,b) show the smoothed non-walking characteristic radius $r_0(i,t)$ for departures during 7:00--7:59 (a) and 17:00--17:59 (b). Dashed concentric circles centered on Hakata Station are shown at $20\,\mathrm{km}$ intervals as distance guides. Panels (c,d) show the corresponding hour-specific weighted attractiveness field $\sum_a\alpha_{a,t}\bigl(n_{i\mid a}/\sum_kn_{k\mid a}\bigr)$ at 7:00 (c) and 17:00 (d).}
  \label{fig:sfig9_fukuoka_prior_diagnostics}
\end{figure}

Figure~\ref{fig:sfig9_fukuoka_prior_diagnostics}, corresponding to Main Fig.~7, shows the non-walking characteristic-radius field and the hour-specific weighted attractiveness field at 7:00 and 17:00. Larger $r_0(i,t)$ values occur in localized station-centered areas and mobility corridors, indicating spatial heterogeneity in the characteristic scale of non-walking mobility. The weighted attractiveness field changes with the activity mixture and is an input to the radiation baseline rather than OD flow itself. Figure~\ref{fig:sfig10_fukuoka_r0_distributions}, corresponding to the Tokyo distributions in Supplementary Fig.~S2, shows that Gaussian smoothing redistributes the pronounced raw near-zero concentration into a more continuous mesoscale distribution and suppresses isolated spikes from weakly supported origin--hour pairs.

\begin{figure}[!htbp]
  \centering
  \includegraphics[width=\linewidth]{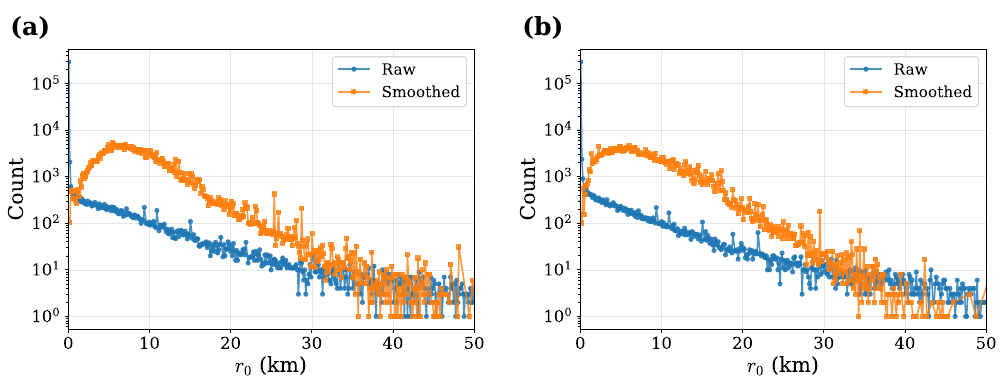}
  \caption{\textbf{Characteristic-radius distributions before and after Gaussian smoothing in Fukuoka.} Panels show the distributions at 7:00 (a) and 17:00 (b), before (raw) and after smoothing with a $300\,\mathrm{m}$ bandwidth. The vertical axis is the binned count of origin cells on a logarithmic scale and the horizontal axis is $r_0$ (km).}
  \label{fig:sfig10_fukuoka_r0_distributions}
\end{figure}

Overall, the Fukuoka analysis reproduces the sequence of Tokyo diagnostics, including spatially adaptive weighting, distance-profile comparisons, controlled-thinning calibration, and characteristic-radius fields. It also shows that excess long-distance mass in the radiation baseline is present in the Fukuoka-wide aggregate, while the refined estimate remains anchored to prominent empirical flows. The Fukuoka analysis uses mode-separated baselines but requires no walking-mode total adjustment.